\documentclass[journal]{IEEEtran}

\ifCLASSOPTIONcompsoc
\usepackage[nocompress]{cite}
\else
\usepackage{cite}
\fi

\usepackage{makeidx,epsfig}
\usepackage{setspace,graphicx,multirow}

\usepackage{amsfonts,latexsym,amssymb,amsthm}

\usepackage{graphicx}
\usepackage{epstopdf}

\usepackage{caption3} 
\DeclareCaptionOption{parskip}[]{} 
\usepackage[small]{caption}

\usepackage{subcaption}

\usepackage{array}

\usepackage[cmex10]{amsmath}

\usepackage{url}

\usepackage{multirow}
\usepackage{hhline}

\usepackage{amsmath}
\usepackage{algorithm}
\usepackage[]{algpseudocode}
\usepackage{varwidth}

\makeatletter
\def\BState{\State\hskip-\ALG@thistlm}
\makeatother

\usepackage{algcompatible}

\usepackage{fixltx2e}

\newcommand*\diff{\mathop{}\!\mathrm{d}}

\usepackage{enumitem}

\makeatletter
  \newcommand\tinyv{\@setfontsize\tinyv{7pt}{9}}
\makeatother

\usepackage{authblk}

\usepackage{textcomp}

\usepackage{dsfont}

\renewcommand\Authands{ and }

\makeatletter
\newcommand{\mathleft}{\@fleqntrue\@mathmargin0pt}
\newcommand{\mathcenter}{\@fleqnfalse}
\makeatother

\usepackage{color}
\ifodd 0
\newcommand{\rev}[1]{{\color{red}#1}} 
\newcommand{\com}[1]{\textbf{\color{blue} (COMMENT: #1)}} 
\else
\newcommand{\rev}[1]{#1}

\newcommand{\com}[1]{}
\fi

\begin{document}
\bibliographystyle{IEEEtran}
\bstctlcite{IEEEexample:BSTcontrol}

\title{Analysis and Optimization for Large-Scale LoRa Networks: Throughput Fairness and Scalability}

\author{Jiangbin~Lyu, ~\textit{Member,~IEEE},
		Dan~Yu,
        and~Liqun~Fu, ~\textit{Senior Member,~IEEE}%
\thanks{This work was presented in part at IEEE ICNC 2020 \cite{LoRaMaxMinArXiv}. This work was supported in part by the National Natural Science Foundation 
	of China (No. 61801408 and No. 61771017), the Natural Science Foundation of Fujian 
	Province (No. 2019J05002), and the Fundamental Research Funds for the Central 
	Universities (No. 20720190008).}%
\thanks{The authors are with the School of Informatics, and the Key Laboratory of Underwater Acoustic Communication and Marine Information Technology (Ministry of Education), Xiamen University, China (email: \{ljb, liqun\}@xmu.edu.cn; yudanx@stu.xmu.edu.cn). \textit{Corresponding author: L. Fu}.}%
\thanks{Copyright \textcopyright 2021 IEEE. Personal use of this material is permitted. However, permission to use this material for any other purposes must be obtained from the IEEE by sending a request to pubs-permissions@ieee.org.}%
}


\markboth{IEEE Internet of Things Journal}%
{To appear}
%



\maketitle

\begin{abstract}
	
LoRa networks are pivotally enabling Long Range connectivity to low-cost and power-constrained user equipments (UEs) in a wide area, whereas a critical issue is to effectively allocate wireless resources to support potentially massive UEs while resolving the prominent near-far fairness issue,
	which is challenging due to the lack of tractable analytical model and the practical requirement for low-complexity and low-overhead design.
	Leveraging on stochastic geometry, especially the Poisson rain model, we derive (semi-) closed-form formulas for the aggregate interference distribution, packet success probability and hence system throughput in both single-cell and multi-cell setups with frequency reuse, by accounting for channel fading, random UE distribution, partial packet overlapping, and/or multi-gateway packet reception.
	The analytical formulas require only average channel statistics and spatial UE distribution, which enable tractable network performance evaluation and incubate our proposed Iterative Balancing (IB) method that quickly yields high-level policies of joint spreading factor (SF) allocation, power control, and duty cycle adjustment for gauging the average max-min UE throughput or supported UE density with rate requirements.
	Numerical results validate the analytical formulas and the effectiveness of our proposed optimization scheme, which greatly alleviates the near-far fairness issue and reduces the spatial power consumption, while significantly improving the cell-edge throughput as well as the spatial (sum) throughput for the majority of UEs, by adapting to the UE/gateway densities.
\end{abstract}

%
\section{Introduction}

%
%
%
%
%
%

\IEEEPARstart{T}{he} Internet of Things (IoT) has found fast-growing applications over recent years in the civilian domain such as for environmental monitoring, building automation and smart cities, which call for wireless technologies that enable low-cost, large-scale, and ultra-durable connectivity for almost everything.
Low Power Wide Area Network (LPWAN)\cite{LPWANoverview} is one of the IoT paradigms that targets at providing long range wireless connectivity to power-constrained IoT devices in a wide area, which includes pronounced technologies such as Narrow Band (NB)-IoT \cite{NBiotMagazine2016} in the licensed band, and LoRa (Long Range)\cite{LoRaWAN} in the unlicensed band.

This paper focuses on LoRa, one of the most promising LPWAN technologies proposed by Semtech\cite{SemTechLoRaModulationBasics} and further promoted by the LoRa Alliance\cite{LoRaWAN}, which attracts a lot of interests from both academia and industry (see e.g., the recent surveys \cite{LPWANoverview} and \cite{LoRaSurvey} and the references therein).
By adaptively trading bit rates for better link budgets, the LoRa physical layer enables flexible long-range communication with low power consumption and low-cost design, which is suitable for those user equipments (UEs) that transmit little amount of data over long periods of time, e.g., water and gas meters.
However, due to the wide coverage area, there are potentially massive UEs to be connected by each LoRa gateway (GW). Moreover, the near-far fairness issue becomes more prominent, as the cell-edge UEs suffer from more severe path-loss and thus more packet failures in the presence of co-channel interference.
Worse still, the UEs' uplink transmissions are typically uncoordinated and may experience random channel fading/packet overlapping. To date, how to model and characterize the performance of such a large-scale LoRa network, and thereby effectively allocate wireless resources to support \textit{massive connectivity with fairness in a wide area}, remains as a challenging and critical issue.

The LoRa physical layer adopts the robust Chirp Spread Spectrum (CSS) modulation with different spreading factors (SFs), where higher SF is associated with larger spreading gain, thus extending the communication range at the cost of lower bit rate.
Moreover, LoRa CSS has a pseudo-orthogonal characteristic with different-SF signals \cite{SemTechLoRaModulationBasics} which can be exploited to accommodate multiple UEs in one channel.
On top of the physical layer, LoRa Alliance has defined the higher layers and network architecture termed as LoRaWAN\cite{LoRaWAN}, whereby the medium access control (MAC) layer is essentially an Aloha variant of random access owing to its simplicity. 
In particular, the class A-type devices in LoRaWAN consume the lowest power as they adopt the pure (unslotted) Aloha-like random access with no synchronization or scheduling overhead, and thus are well suited for low duty-cycle\footnote{The duty cycle reflects the time intensity of random access, which can be regulated by the network server for traffic shaping\cite{LoRaWAN}.} devices which are idle most of the time.
Therefore, we consider class A type with pure Aloha in this paper, which is the simplest and also mandatory for all LoRa devices to implement.\footnote{Note that further performance improvement can be achieved by means of light-weight scheduling and coarse\cite{LoRaIoTJ2018Pollin} or complete\cite{LoRaIoTJ2019LowOverhead} synchronization, which is left for future work.}



Despite the low duty cycle, in wide-area or dense deployment scenarios, LoRa networks will still suffer from collisions of concurrent transmissions in the same channel and SF. 
The conventional protocol model\footnote{Two packets are considered both failure if they overlap by any part in time.} for pure Aloha does not account for the channel capture effect\footnote{A strong-enough packet could reject the co-channel interference and be correctly received, a.k.a. \textit{channel capture}. Details are provided in Section \ref{SectionModel}.} that depends on transmit power (TP), channel fading, and aggregate interference which in turn depends on both spatial randomness of UE distribution and time randomness of (partial) packet overlapping.
A simulation model based on real interference measurements is presented in \cite{LoRaSensors2017}, while a scalability analysis of LoRa networks is performed in \cite{LoRaNS3IoTJ} using a LoRa error model together with the LoRaWAN MAC protocol in the ns-3 simulator.
In addition, measurement trials (e.g., \cite{SemtechDenseTrail}) are carried out to justify the practical deployment of LoRa networks.
However, as field trials are costly and system-level simulations are time-consuming, the investigated network is typically limited in size and/or lack flexibility for quick and fine-tuned analysis.
To this end, a \textit{model-based methodology} is adopted in this paper, by seeking tractable analytical models for the large-scale LoRa network, based on which general network analysis/capacity planning can then be performed to reveal insights on network scalability and thereby provide general deployment suggestions.

Along this line of research, stochastic geometry has been extensively applied for tractable modeling/analysis of the spatial/temporal randomness of wireless networks\cite{AndrewsCellular}\cite{SpatialAlohaSlotted}.
In the context of LoRa networks, scalability analysis is conducted in \cite{LoRaWCL2017Raza} and \cite{MahmoodInterSF} via stochastic geometry, which neglects the time dependence of (partially) overlapping packets.
Accurate packet overlapping is considered in \cite{LoRaCL2018Korean} and \cite{2DinterferenceICC2017}, and yet it is difficult to obtain the exact distribution of packet success probability analytically.
For analytical tractability, we hence leverage another thread of stochastic geometry for \textit{non-slotted} Aloha, i.e., the Poisson rain model\cite{NonSlottedINFOCOM2010}, which caters for channel fading, aggregate interference, and accurate packet overlapping. 
The Poisson rain model is recently applied in \cite{PoissonRainLoRa} to analyze a single-cell LoRa network whereby the SFs are tuned to be inversely related to the respective receiver sensitivity in order to equalize the packet reception probability of all UEs.
However, the above work has not considered throughput fairness and the options of TP and duty cycle control, thus leaving an open question on the network scalability in the general multi-cell setup under jointly optimized design.








Motivated by the above, to achieve massive connectivity (in terms of throughput per UE) with fairness, we study the problem of maximizing the minimum throughput of all UEs in both the single-cell and multi-cell setups,
by jointly optimizing the SF, TP, and duty cycle control.
The formulated problem can be shown to be a mixed-integer non-linear programming (MINLP) due to the discrete SF allocation and nonconvex throughput constraint, which is thus challenging to solve.
To this end, our main contributions are summarized as follows.
\begin{itemize}[leftmargin=0.14in]
	\item By leveraging the \textit{Poisson rain model} and average channel statistics, (semi-) closed-form formulas are derived for the aggregate interference distribution, packet success probability and hence system throughput in both single- and multi-cell setups, by accounting for channel fading, power control, random UE distribution, partial packet overlapping, and/or multi-GW packet reception.

	\item Based on the derived analytical results, a novel \textit{Iterative Balancing (IB) method} is further proposed which quickly yields high-level policies of joint SF, TP and duty cycle control for gauging the average max-min UE throughput or supported UE density with rate requirements.
	
	\item Analytical formulas for approximating the \textit{optimal duty cycle} within each SF group are derived, which provide quick reference for traffic shaping under given UE/GW densities.
	
	\item Different multi-cell frequency reuse schemes are investigated which trade off throughput performance with the limited number of concurrent demodulation paths in practical LoRa GWs. Moreover, a modified fractional frequency reuse scheme, termed as \textit{LoRa-FFR}, is proposed by accounting for the pseudo-orthogonal SFs.
	
	\item Simulation results validate the analytical formulas and the effectiveness of our proposed optimization scheme, which greatly alleviates the near-far fairness issue and reduces the spatial power consumption, while significantly improving the cell-edge throughput as well as the spatial (sum) throughput for the majority of UEs, by adapting to the UE/GW densities.	
\end{itemize}

Finally, we briefly compare our proposed method with other existing works.
An Adaptive Data Rate (ADR) mechanism is recommended by Semtech\cite{SemTechADR} to automatically adapt each UE's SF and TP based on individual link budget. However, simulation studies in \cite{RazaADR} reveal that if link conditions change or network size becomes too large, the convergence time of the ADR mechanism to a desired communication setting is quite long. 
Various improved ADR schemes are thus proposed to dynamically optimize the data rate, airtime, and energy consumption (see the recent survey\cite{ADRsurvey} and the references therein).
In particular, instead of rate adaptation merely based on 
link-level budget/measurements\cite{SemTechADR}, 
a better strategy is the system-level load balancing that exploits the quasi-orthogonality and diverse communication range of different SFs\cite{BianchiLoRaTWC,LoRaPollinICC2017,LoRaWCL2017Raza,LoRaCL2018Korean,MahmoodInterSF,HeusseSpatial,QinZhijinEE}, which is also adopted in this paper.
In \cite{BianchiLoRaTWC}, SFs are assigned by equalizing the time-on-air of packets sent by each UE, while in \cite{LoRaPollinICC2017} the optimal proportion of SFs is calculated to assist the SF and TP allocation by minimizing the collision probability within each SF group. However, these works\cite{BianchiLoRaTWC}\cite{LoRaPollinICC2017} focus more on the protocol design and rely on simplified channel contention models.
Another widely adopted SF allocation strategy is to 
assign higher SF to farther UEs based on their distances to the GW, whereby the distance thresholds are set according to different rules such as equal-interval-based\cite{LoRaWCL2017Raza}, equal-area-based\cite{LoRaCL2018Korean}, 
signal-to-noise-ratio (SNR)-based\cite{MahmoodInterSF}, and packet-delivery-ratio (PDR)-based\cite{LoRaCL2018Korean}\cite{HeusseSpatial}, which rely on distance/path-loss estimations based on statistical channel state information (CSI). 
In the case when instantaneous CSI is available, the network performance such as energy efficiency\cite{QinZhijinEE}
can be further improved by solving sophisticated problems (typically belonging to MINLP due to discrete channel/SF allocation and/or user scheduling) with advanced optimization techniques, at the cost of increased complexity and communication overhead.
Different from the above works\cite{LoRaWCL2017Raza,LoRaCL2018Korean,MahmoodInterSF,HeusseSpatial,QinZhijinEE}, our proposed IB method builds upon the tractable Poisson rain model, and aims at achieving the average max-min throughput in large-scale LoRa networks under given network parameters.

The rest of this paper is organized as follows.
The system model is introduced in Section \ref{SectionModel}, while the max-min throughput problem is formulated in Section \ref{SectionFormulation}.
Next, the proposed solution for the single-GW LoRa network is presented in Section \ref{SectionSingle}, which is extended to the multi-GW scenario under different frequency reuse schemes in Section \ref{SectionMultiGW}.
Further extension to account for macro diversity of multi-GW reception is discussed in Section \ref{SectionMultiGWreception}.
Numerical results are provided in Section \ref{SectionSimulation}.
Finally, we conclude the paper in Section \ref{SectionConclusion}.

\section{System Model}\label{SectionModel}

\subsection{Network Model}
\begin{figure}
	\centering
	\hspace{-10pt}
	\begin{subfigure}[b]{0.5\linewidth}
		\includegraphics[width=1\linewidth,  trim=0 0 0 0,clip]{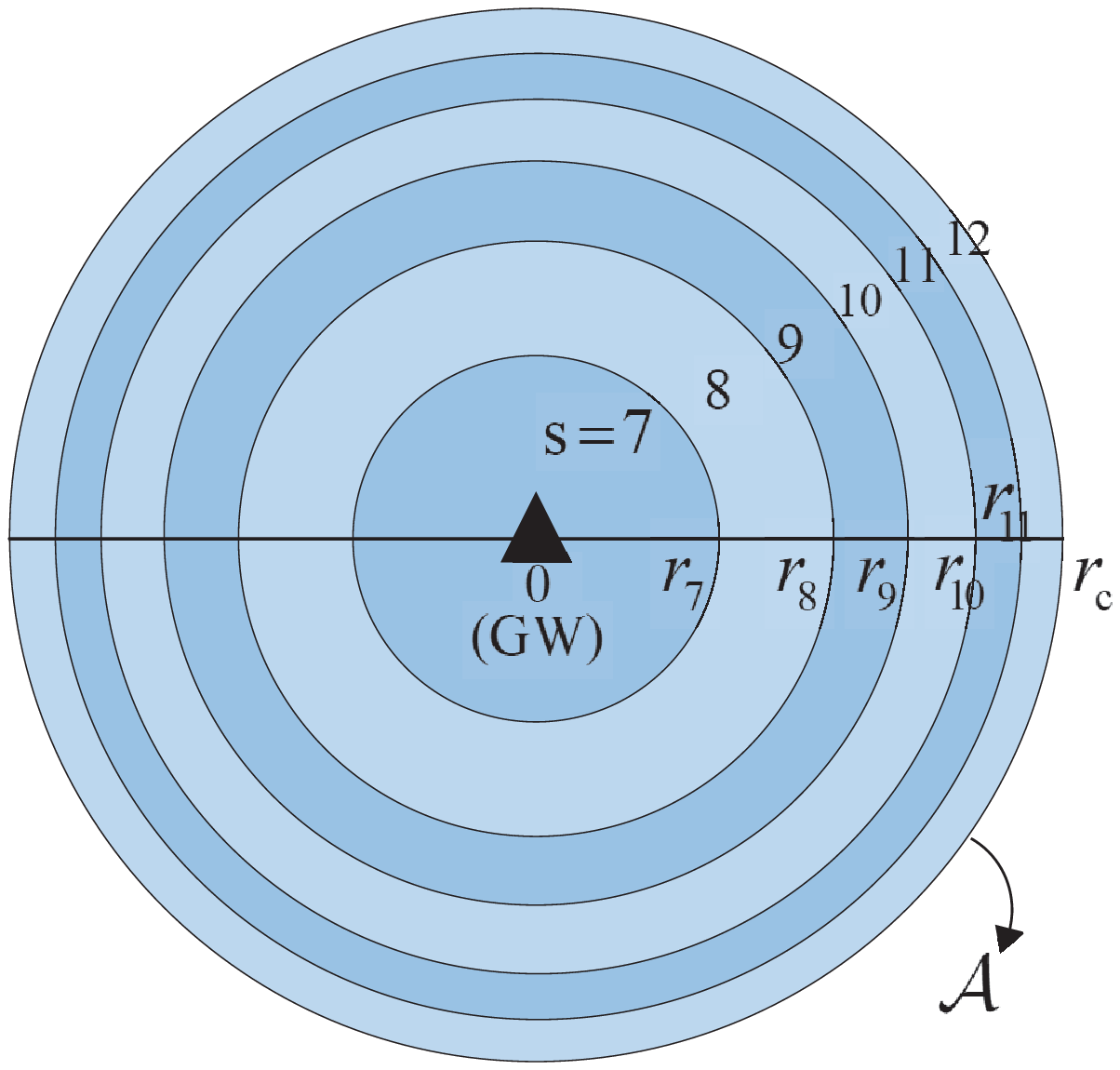}
		\caption{\vspace{0ex}}\label{LoRa1}
	\end{subfigure}%
	\begin{subfigure}[b]{0.5\linewidth}
		\includegraphics[width=1\linewidth,  trim=280 50 260 0,clip]{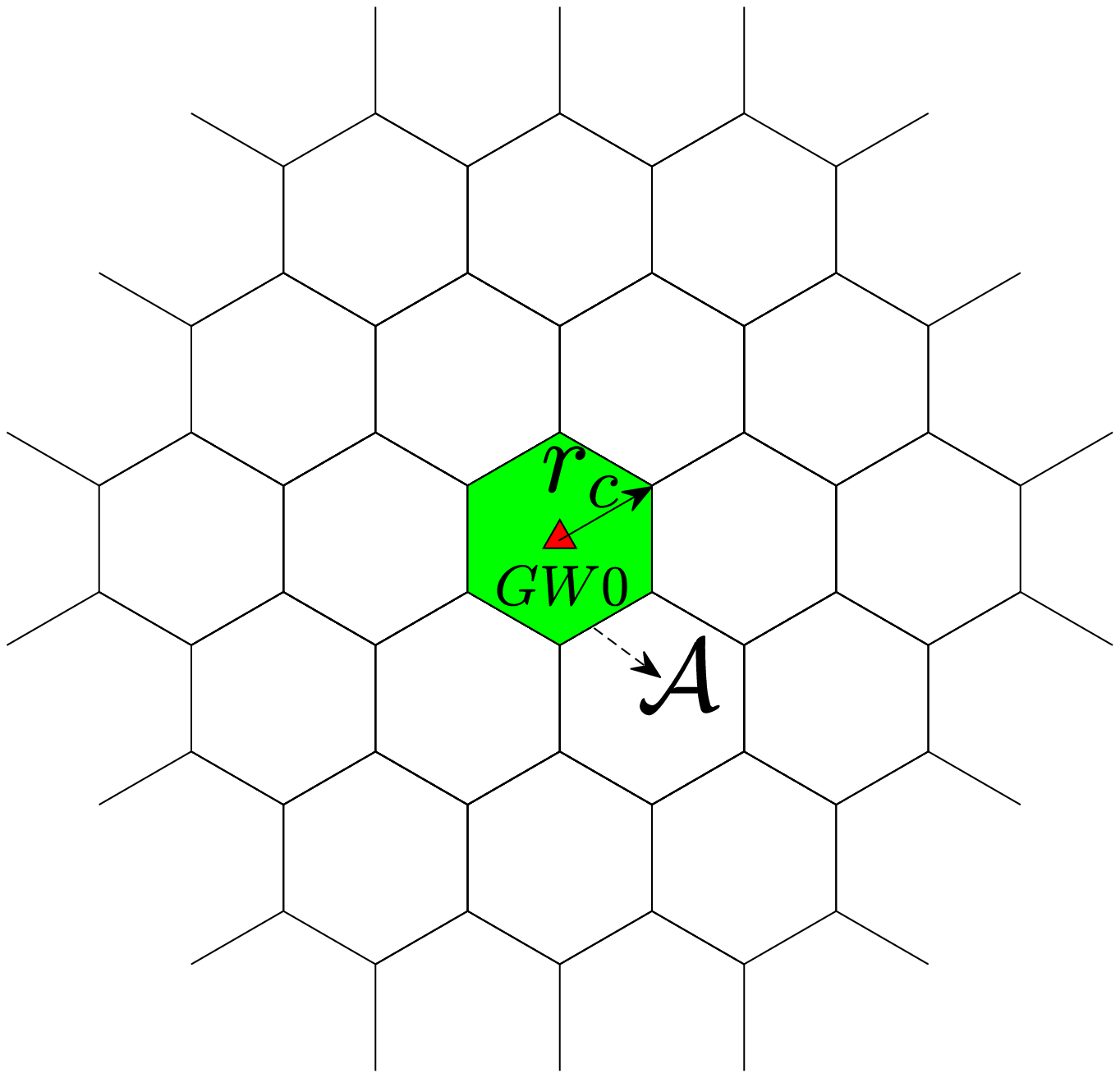}
		\caption{\vspace{0ex}}\label{LoRa2}
	\end{subfigure}%
	\caption{(a) Single-GW LoRa network. (b) Hexagonal cell layout in multi-GW LoRa networks.\vspace{-2ex}}\label{LoRa}
\end{figure}

Consider uplink communication\footnote{Uplink communication from LoRa end devices to the network server is typically uncoordinated and presents as the performance bottleneck. Similar to other prior works \cite{LoRaWCL2017Raza,MahmoodInterSF,LoRaCL2018Korean,2DinterferenceICC2017,PoissonRainLoRa,BianchiLoRaTWC,LoRaPollinICC2017} that focus on uplink scalability, we consider unconfirmed traffic that consists of only uplink data packets.} from distributed UEs to LoRa GWs.
Consider a typical GW 0 located at the origin which is also the center of the typical cell 0 with a two-dimensional (2D) region $\mathcal{A}\subset\mathbb{R}^2$ of radius $r_c$ meters (m).
The typical cell region $\mathcal{A}$ is considered as a disk region in the single-GW scenario as shown in Fig. \ref{LoRa}(a).
In the multiple-GW scenario, the classic hexagonal grid cell layout is considered,\footnote{The regular grid-based GW layout serves as a baseline for evaluating the network performance under planned topology\cite{BianchiLoRaTWC,LoRaIoTJ2018Pollin,LoRaNS3IoTJ}, while tailored GW planning\cite{JimingLoRa} for specific LoRa networks can bring further performance improvement and is left for future work.}
where each GW is at the center of its hexagonal cell with radius $r_c$ and inter-cell distance of $\sqrt{3}r_c$, as shown in Fig. \ref{LoRa}(b).
Moreover, consider $M$ tiers of other GWs denoted by the sets $\mathcal{M}_m, m=1,2,\cdots,M$, where the GWs in tier $m$ have the same distance $D_m$ from GW 0. By default, denote $m=0$ as the 0-th tier which consists of GW 0 only.
Further denote $\mathcal{M}\triangleq \bigcup_{m=0}^M \mathcal{M}_m$ as the set of all GWs, $\bold v_n\in \mathbb{R}^2$ as the horizontal location of GW $n\in\mathcal{M}$, and $H_G$ as the GW height.


Different from the conventional cellular network, a LoRa uplink packet can be potentially received by multiple nearby GWs to increase the packet success probability. The uplink transmission is successful if the packet can be decoded by one of the GWs, which then further forwards the packet to the network server (NS) via wired backhaul.\footnote{The NS can be equipped with higher computational capability than GWs and thus can handle more sophisticated tasks including ADR operations.}
For the purpose of exposition, we first focus on the uplink reception by GW 0 in Sections \ref{SectionModel}, \ref{SectionFormulation}, \ref{SectionSingle} and \ref{SectionMultiGW}, and then extend to the general setup with multi-GW reception in Section \ref{SectionMultiGWreception}.


Consider a given frequency reuse factor of $\omega$ ($0<\omega\leq 1$) for the multiple GWs to share a set of $\mathrm{F}$ orthogonal channels each with bandwidth $B$ Hz.
As a baseline scheme, consider full frequency reuse ($\omega=1$) where each GW uses the full set of channels as in the state-of-the-art LoRaWAN solutions.\footnote{Other frequency reuse schemes will be discussed later in Section \ref{SectionMultiGW}, which trade off throughput performance with the limited number of concurrent demodulation paths in practical LoRa GWs, thereby providing possible options of network planning for future release of LoRaWAN solutions.} 
Assume that each UE randomly selects its operating channel, and as a result the UEs are uniformly distributed over the $\mathrm{F}$ channels.
For the purpose of exposition, we focus on one typical channel to characterize the co-channel interference and hence throughput performance.
In the whole considered region and on one typical channel,
denote $\mathcal{K}=\{1,2,\cdots,K\}$ as the set of UEs which have packets to transmit in the considered time period, with 2D locations $\bold w_k\triangleq (x_k,y_k),x_k,y_k\in\mathbb{R}, k\in\mathcal{K}$ on the ground, which are assumed to
follow a homogeneous\footnote{Our analytical framework can be extended to handle the inhomogeneous PPP case with non-homogeneous intensity measure\cite{HeusseSpatial} in the considered area. Nevertheless, the HPPP assumption serves as a good baseline for evaluating the general network performance.} Poisson point process (HPPP) $\Phi\subset \mathbb{R}^2$ with density $\lambda$ /m$^2$.

Note that in the multi-GW scenario, the set $\mathcal{K}$ can be further divided into $M+1$ disjoint subsets $\mathcal{K}^{(m)}$, $m=0,1,\cdots,M$ where $\mathcal{K}^{(m)}$ denotes the set of UEs in tier-$m$ cells. In the rest of the paper, we use the superscript $(m)$ to mark a variable associated with the tier-$m$ cell.

\subsection{Channel Model}


LoRa is adopted as the physical layer transmission technology for UE-GW communications.
For the purpose of exposition, consider one of the channels with pseudo-orthogonal SFs $s\in\mathcal{S}\triangleq\{7,\cdots,12\}$.
For a given $s$, the transmission rate in bits/second (bps) is given by\cite{SemTechLoRaModulationBasics}
\begin{equation}
R_s=\frac{s}{2^s}BC,
\end{equation}
where $C$ is the code rate.
For simplicity, assume that the packets are of equal length $L_s$ bits for a given $s$, which corresponds to a packet duration of $T_s\triangleq L_s/R_s$.

Denote $P_k$ as the TP of UE $k$ in Watt (W), capped by the maximum allowed value $P_\textrm{max}$.
Assume that the GWs and UEs are each equipped with a single omnidirectional antenna with unit gain.
We assume a simplified fading channel model without shadowing, which consists of distance-dependent path-loss with path-loss exponent $n_0\geq 2$ and an additional random term $\zeta$ accounting for small-scale fading.
Note that shadowing effect can also be incorporated into the stochastic geometry based analytical framework by treating it as equivalent random perturbation in the locations of the transmitters (see Sections III.G and VI.A in \cite{AndrewsPrimer} and the references therein), which is ignored in this work for simplicity.
Therefore, the channel power gain from UE $k$ to GW 0 can be modeled as
$g_k=\bar g_k\zeta_k$, where $\bar g_k\triangleq\alpha_0 (H_G^2+d_k^2)^{-n_0/2}$ is the average channel power, with $\alpha_0=(\frac{4\pi f_c}{c})^{-2}$ denoting the average channel power at a reference distance of 1 m, $f_c$ denoting the carrier frequency, $c$ denoting the speed of light, and $d_k$ denoting the horizontal distance from UE $k$ to GW 0;
and $\zeta_k\stackrel{\textrm{dist.}}{=}\zeta\sim \textrm{Exp}(1)$ is an independent and identically distributed (i.i.d.) exponential RV with mean $1$ accounting for the small-scale Rayleigh fading,\footnote{The proposed analytical method in this paper can be extended to account for other fading channels such as the more general Nakagami-$m$ fading\cite{AndrewsGeneralFading}.} with $\stackrel{\textrm{dist.}}{=}$ denoting equal in distribution.

\subsection{SNR and SIR}

Consider a typical UE 0 with SF $s$ to communicate with GW 0.
Assume that the receiver noise at GW 0 is additive white Gaussian noise (AWGN) with zero mean and power $\sigma^2$.
In the case without interference, the received signal-to-noise ratio (SNR) is given by
\begin{equation}
\eta_{s,0}\triangleq  P_0 g_0/\sigma^2= P_0\bar g_0\zeta_0/\sigma^2,
\end{equation}
which needs to be not smaller than a certain threshold $\bar\eta_{s}$ in order for the packet to be successfully decoded,
where $\bar\eta_{s}$ is typically lower for higher SF (see, e.g., Table I in \cite{LoRaWCL2017Raza}).

On the other hand, when multiple UEs transmit concurrently on the same channel,
it is possible for the reference packet to reject the co-channel interference and be correctly received (a.k.a. \textit{channel capture}) if its signal-to-interference ratio (SIR) is a few dBs in case of co-SF interference (6 dB as in \cite{LoRaThreshold2015}, or 1 dB as in \cite{CroceCLinterSF}) or as low as $-8 \sim -25$ dB in case of inter-SF interference depending on the interfering SF\cite{CroceCLinterSF}.
First, we introduce the case with co-SF interference only.
Consider unslotted Aloha as the MAC-layer multiple access method, which corresponds to Class A in LoRaWAN.
Denote $\mathcal{K}_s\subseteq \mathcal{K}$ as the set of UEs with SF $s$, i.e.,
\begin{equation}
\mathcal{K}_s=\{k|s_k=s, k\in\mathcal{K}\},
\end{equation}
where $s_k$ is the SF of UE $k$.
Denote the duty cycle, i.e., fraction of time that a UE is transmitting packets, as $\delta_k\in [0,1]$ for UE $k$, which is subject to a certain limit $\Delta_\textrm{max}$ (e.g., 1\%) and can be regulated by the network server for traffic shaping\cite{LoRaWAN}.
For simplicity, assume that the UEs in $\mathcal{K}_s$ comply with the same duty cycle level $\delta_k=\Delta_s, \forall k\in \mathcal{K}_s$.
The number of transmission initiations per UE per unit time is then given by
\begin{equation}\label{rho_s}
\rho_s\triangleq \frac{1}{T_s/\Delta_s-T_s}=\frac{\Delta_s}{(1-\Delta_s)T_s},
\end{equation}
which corresponds to the time frequency of channel access.



The aggregate co-SF interference is modeled in the following.
Consider the typical UE 0 with SF $s$ which transmits a reference packet during the time interval $[0,T_s]$.
The average number of co-SF packets initiated during this packet interval is given by $\rho_s T_s |\mathcal{K}_s|$.
At a given time instant $t\in [0,T_s]$, the transmitting UEs with SF $s$ make up a set $\mathcal{K}_s(t)\subseteq\mathcal{K}_s$, which cause the total received interference power at GW 0 as 
\begin{equation}\label{Ist}
I_s(t)\triangleq \sum_{k\in\mathcal{K}_s(t)\setminus\{0\}} P_k g_k=\sum_{k\in\mathcal{K}_s(t)\setminus\{0\}} P_k \bar g_k \zeta_k.
\end{equation}

Second, 
it is also possible to incorporate inter-SF interference in our analytical framework by introducing the \textit{cross-correlation factor} $\chi_{s,s'}\in[0,1)$ between the target SF $s$ and other SF $s'$, which represents the cross-correlation of interfering energy between these two different spreading codes.
In this case, the co-channel interference at time $t$ for the typical UE 0 with SF $s$ is given by
\begin{equation}\label{IstInterSF}
I_s(t)\triangleq \sum_{k\in\mathcal{K}_s(t)\setminus\{0\}} P_k \bar g_k \zeta_k +\sum_{s'\in\mathcal{S}\setminus\{s\}} \chi_{s,s'}\bigg(\sum_{k\in\mathcal{K}_{s'}(t)} P_k \bar g_k \zeta_k\bigg),
\end{equation}
where $\mathcal{K}_{s'}(t)$ denotes the set of transmitting UEs with SF $s'$ at time $t$.

The aggregate interference power at GW 0 averaged over one packet duration $T_s$ is then given by
\begin{equation}\label{barIs}
\bar I_s\triangleq \frac{1}{T_s}\int_{0}^{T_s} I_s(t) \diff t.
\end{equation}
Under the pure Aloha model with average interference constraint\cite{NonSlottedINFOCOM2010}, 
the SIR $\gamma_{s,0}$ of the reference packet is given by
\begin{equation}\label{SINR0}
\gamma_{s,0}\triangleq P_0\bar g_0\zeta_0/\bar I_s,
\end{equation}
which corresponds to the situation where some coding with repetition and interleaving of bits on the whole packet duration is used; and 
the reference packet can be successfully decoded if its SIR is not smaller than a certain threshold $\bar\gamma_s$ (see, e.g., Table 1 in \cite{LoRaThreshold2015}).
Finally, note that the formulas \eqref{barIs} and \eqref{SINR0} are applicable to both the case with co-SF interference only in \eqref{Ist} and the case with inter-SF interference as well in \eqref{IstInterSF}.
Nevertheless, since different SFs are pseudo-orthogonal and the co-channel interference comes mostly from co-SF UEs especially under power control\cite{LoRaPollinICC2017}\cite{HeusseSingleCell}, 
we ignore the inter-SF interference in the rest of the paper for simplicity, and focus on characterizing the achievable capacity upper bound under joint SF/TP/duty cycle control.

\subsection{Packet Success Probability and Throughput}


Due to the unique characteristics of LoRa CSS modulation\cite{LoRaThreshold2015,CroceCLinterSF}, a success event of the LoRa packet needs to satisfy both the SNR and SIR conditions mentioned above\cite{LoRaThreshold2015,HeusseSingleCell,BianchiLoRaTWC,LoRaWCL2017Raza,LoRaCL2018Korean}.\footnote{Note that our analysis can be readily extended to the case where a threshold of the signal-to-interference-plus-noise ratio (SINR) for successful LoRa packet transmission is available.} 
The success probability of a reference packet sent by UE 0 with SF $s$ and received by GW 0 is then given by
\begin{align}
\textrm{P}_{s,0}^{\textrm{suc}}&\triangleq\mathbb{P}\big\{\{\eta_{s,0}\geq\bar\eta_s\} \&\{\gamma_{s,0}\geq\bar\gamma_s\}\big\}\label{PsusTrue}\\
&=\mathbb{P}\big\{\{P_0\bar g_0\zeta_0/\sigma^2\geq\bar\eta_s\} \&\{P_0\bar g_0\zeta_0/\bar I_s\geq\bar\gamma_s\}\big\}\notag\\
&=\mathbb{P}\big\{\{\zeta_0\geq\bar\eta_s\sigma^2/(P_0\bar g_0)\} \&\{\zeta_0\geq\bar\gamma_s\bar I_s/(P_0\bar g_0)\}\big\}\notag\\
&=\mathbb{P}\big\{\{\zeta_0\geq a\} \&\{\zeta_0\geq b\}\big\}=\mathbb{P}\{\zeta_0\geq a\} \mathbb{P}\{\zeta_0\geq b|\zeta_0\geq a\}\notag\\
&\stackrel{(1^\circ)}{\geq} \mathbb{P}\{\zeta_0\geq a\}\mathbb{P}\{\zeta_0\geq b\}\notag\\
&\stackrel{(2^\circ)}{=}e^{-\bar\eta_s\sigma^2/(P_0\bar g_0)}\mathcal{L}_{\bar I_s}\big(\bar\gamma_s/(P_0\bar g_0)\big),\label{Psuc0}
\end{align}
where $a\triangleq \bar\eta_s\sigma^2/(P_0\bar g_0)$, $b\triangleq \bar\gamma_s\bar I_s/(P_0\bar g_0)$ and $\mathbb{P}\{\cdot\}$ denotes the probability of an event;
$(1^\circ)$ is due to $\mathbb{P}\{\zeta_0\geq b|\zeta_0\geq a\}\geq \mathbb{P}\{\zeta_0\geq b\}$ since the event of $\zeta_0\geq a$ implies that $\zeta_0\geq b$ is more likely to happen;
$(2^\circ)$ is due to $\zeta_0$ with exponential distribution; 
and the term $\mathbb{P}\{\zeta_0\geq a\}=e^{-\bar\eta_s\sigma^2/(P_0\bar g_0)}$ represents the packet success probability based on the SNR condition, while the term $\mathbb{P}\{\zeta_0\geq b\}=\mathcal{L}_{\bar I_s}\big(\bar\gamma_s/(P_0\bar g_0)\big)$ represents that based on SIR, where $\mathcal{L}_{\bar I_s}(\cdot)$ is the Laplace transform\footnote{The Laplace transform of an RV $X$ is defined as $\mathcal{L}_{X}(z)\triangleq \mathbb{E}_{X}\big\{e^{-zX}\big\}$.} of the RV $\bar I_s$.

Note that the inequality $(1^\circ)$ is tight when $b\gg a$ (i.e., $\bar I_s\gg \bar\eta_s\sigma^2/\bar\gamma_s$), which is typically the case for interference-limited scenario, especially thanks to the low value of $\bar\eta_s$ (e.g., -6 dB to -20 dB) associated with the low sensitivity of LoRa receiver\cite{LoRaWCL2017Raza}.
In the rest of the paper, we use the lower bound in \eqref{Psuc0} directly to calculate $\textrm{P}_{s,0}^{\textrm{suc}}$. To this end, we need to characterize the Laplace transform of the average interference $\bar I_s$ in order to obtain $\textrm{P}_{s,0}^{\textrm{suc}}$, and further obtain the throughput of UE 0 in bps given by
\begin{equation}\label{theta0}
\theta_{s,0}\triangleq R_s\Delta_s\textrm{P}_{s,0}^{\textrm{suc}}.
\end{equation}
The throughput of other UE $k$ with SF $s$, denoted as $\theta_{s,k}$, can be obtained similar to $\theta_{s,0}$.

\section{Problem Formulation}\label{SectionFormulation}

In this section, we formulate the optimization problem to maximize the minimum throughput $\theta$ of all UEs by jointly optimizing the TP $P_k$, SF $s_k$, and duty cycle $\Delta_s$ of all UEs $k\in\mathcal{K}$, under given cell radius $r_c$, UE density $\lambda$ per channel, and a certain frequency reuse pattern with reuse factor $\omega$.
The problem can be formulated as
\begin{align}
\mathrm{(P1)}: \underset{
\begin{subarray}{c}
  \theta,\Delta_s, s\in\mathcal{S}\\
  P_k, s_k, k\in\mathcal{K}
  \end{subarray}
}{\max}& \quad\theta \notag\\
             \text{s.t.}\quad&\theta_{s,k}\geq \theta,\quad \forall k\in\mathcal{K}_s, s\in\mathcal{S},\label{ConstraintTheta}\\ 
             &0\leq \Delta_s\leq \Delta_\textrm{max},\quad s\in\mathcal{S},\label{ConstraintDelta}\\            
             &0\leq P_k\leq P_\textrm{max},\quad k\in\mathcal{K},\label{ConstraintP}\\ 
             &s_k\in\mathcal{S},\quad k\in\mathcal{K}.\label{ConstraintS}       
\end{align}

The problem (P1) is a MINLP due to the discrete SF allocation and the nonlinear/nonconvex constraint \eqref{ConstraintTheta}.
In general, (P1) requires $O(|\mathcal{S}|^{K})$ complexity to exhaustively search for the optimal SF allocation, and it is still a nonconvex problem with $O(K)$ variables even for fixed SF allocation, which is thus prohibitive to solve for the scenario with massive IoT devices.

In addition to the complexity, solving (P1) also faces practical challenges.
Due to the low energy consumption requirement, the UEs typically have low wake-up frequency and limited active time upon wake-up. 
Therefore, the set of active UEs may be constantly changing over time from the pool of massive IoT devices.
As a result, it would be difficult for the GW(s) to obtain the instantaneous CSI of all UEs at all time which is required to solve (P1) optimally.
Moreover, the GW(s) has limited downlink capacity to send control or feedback information to UEs.
Therefore, even if the centralized problem (P1) can be solved instantaneously for every given snapshot of active UEs, it would cause a huge overhead for GW-UE handshaking and conveying the optimal solutions to each UE.


In order to resolve the above challenges, 
we propose a different approach by first characterizing the statistical distribution of the aggregate interference and hence the packet success probability and throughput averaged over a larger time scale, instead of detailed optimization on a short time basis.
Specifically, instead of requiring detailed information about instantaneous CSI at a given time snapshot, our proposed analytical framework requires only knowledge about the active UE distribution and density over a certain period of time as well as the channel statistics such as fading distribution and path-loss exponent, thus allowing for tractable and quick network performance evaluation.
We seek to maximize the minimum throughput averaged over time, by jointly designing \textit{high-level policies} of SF/TP/duty cycle control that adapt to network parameters including UE/GW densities.



\textit{Remark 1:}
The average max-min throughput can be translated into an overall estimate of the maximum supported UE density with rate requirements under given GW density and other network parameters, which provides useful and quick reference for network planning to satisfy the overall throughput demand of large-scale LoRa networks. Such network capacity planning is needed especially for massive Machine Type Communications (mMTC) services due to the massive number of devices involved, which are more likely to cause network congestions.
In addition, the max-min fairness serves as a baseline to evaluate the network performance when the data from all users (e.g., sensors) have the same value/priority. On the other hand, for the user data with heterogeneous value/priority, other types of fairness (e.g., proportional fairness that assigns resource based on priority) can be applied so that the data with lower priority can still have a chance to get through instead of being blocked all the time, which is left for future investigation.

\textit{Remark 2:}
In practice, the knowledge of active user distribution and density can be estimated based on collected location information during the deployment phase along with the user traffic/mobility information collected in the current time period, while the knowledge of channel statistics can be estimated according to the type of environment and/or channel measurement campaigns in the site of interest, which is worth the effort for planning large-scale network deployment. Note that a complete treatment for estimating network parameters is beyond the scope of this paper. Nevertheless, in the case with potential bias of the network knowledge, our proposed iterative balancing method (discussed later in Section \ref{SectionSingle}) can also be applied, by first obtaining a good initial operating point based on the analytical model and estimated network parameters, and then fine-tuning the SF/TP control based on the measured average throughput of the UEs. For simplicity in this paper, we focus on the network analysis/planning under given network parameters as is typically assumed in the stochastic geometry method, which has been extensively adopted in the wireless communication research literature for tractable network analysis, including recent works for LoRa networks such as \cite{LoRaWCL2017Raza,MahmoodInterSF,LoRaCL2018Korean,2DinterferenceICC2017} and \cite{PoissonRainLoRa}.

\section{Proposed Scheme for Single-GW LoRa Network}\label{SectionSingle}

In this section, we consider the single-GW scenario as in Fig. \ref{LoRa}(a), with the overall UE set $\mathcal{K}=\mathcal{K}^{(0)}$ representing the UEs in the single disk cell area $\mathcal{A}$.\footnote{For brevity, throughout this section, we use the superscript $(0)$ to mark a variable associated with the considered typical cell 0, without further definition.}
In general, higher SF is associated with lower SNR threshold $\bar\eta_s$ at the cost of lower data rate\cite{SemTechLoRaModulationBasics}, which helps to extend the communication range from the GW in the absence of interference.
Therefore, it is natural to assign higher SF to UEs at a longer distance from the GW, which have larger path-loss and hence lower average received power at the GW under the same TP.
We thus adopt the distance-based SF allocation policy which can be implemented by using distance/path-loss estimations based on statistical CSI instead of instantaneous CSI.


Specifically, for the considered cell area $\mathcal{A}$ of radius $r_c$, it is partitioned into $|\mathcal{S}|=6$ zones with the delimiting distance threshold $r_s, s=7,\cdots,11$. By default, denote $r_6=0$ and $r_{12}=r_c$. The UEs in each zone $\mathcal{A}_s$ within distance $r_{s}$ to $r_{s-1}$ are allocated with the SF $s$, respectively.\footnote{Note that in practical implementations, the distance thresholds can be translated into equivalent path-loss thresholds. Our analysis can be extended to account for the shadowing effect, whereby the sharp distance boundaries for SF allocation could be replaced by ``soft" boundaries with overlapping regions between different SF zones, as suggested by \cite{MahmoodInterSF}.}
For example, the UEs within distance $r_7$ from GW 0 are allocated with SF 7, the UEs within distance $r_8$ to $r_7$ are allocated with SF 8 and so on, as shown in Fig. \ref{LoRa}.
Denote $\bold r\triangleq (r_7,\cdots,r_{11})$ as the vector of partitioning distance threshold, which is the optimization variable to determine the SF allocation.
For HPPP distributed UEs, the number of UEs $K_s^{(0)}$ with SF $s$ in cell 0 is a Poisson RV with mean $\lambda A_s$, where $A_s=|\mathcal{A}_s|$ denotes the area in cell 0 associated with SF $s$ and is given by
\begin{equation}\label{Area}
A_s=\pi (r_s^2-r_{s-1}^2), s=7,\cdots,12.
\end{equation}

When the partitioning distance threshold $\bold r$ is given, the power control problem is also simplified.
If the typical UE is at the zone edge with distance $r_s$ from GW 0, its packet success probability follows from \eqref{Psuc0}, and thus is more likely to be in outage due to larger path-loss than the inner UEs with the same SF $s$, under the same TP.
In order to achieve the average max-min throughput for the UEs $k\in\mathcal{K}_s$, we propose the ``slow" channel inversion power control based on the average channel power gain $\bar g_k$, such that the average received power at GW 0 is the same\footnote{For simplicity, continuous power control is considered here while the obtained results can be quantized into discrete power levels for implementation.} for all UEs with SF $s$, denoted by $\bar Q_{s}^{(0)}$.
Specifically, the TP of each UE $k\in\mathcal{K}_s$ is given by
\begin{equation}\label{PowerControlk}
P_k=\bar Q_{s}^{(0)}/\bar g_k=\bar Q_{s}^{(0)}(H_G^2+d_k^2)^{n_0/2}/\alpha_0, \forall k\in\mathcal{K}_s,
\end{equation}
where the TP $P_k$ is inversely proportional to $\bar g_k$, and thus we have $\bar Q_{s}^{(0)}=P_k\bar g_k, \forall k\in\mathcal{K}_s$, including the typical UE 0.
The power control in \eqref{PowerControlk} can also be written in the form of distance-based policy as follows:
\begin{equation}
P(s,r)\triangleq \bar Q_{s}^{(0)}(H_G^2+r^2)^{n_0/2}/\alpha_0,
\end{equation}
where $P(s,r)$ denotes the TP of the UE with SF $s$ at distance $r$ from its receiving GW.
In particular, for the UE at the zone edge with TP $P_s^\textrm{edge}\triangleq P(s,r_s)$, we have 
\begin{equation}\label{Qs}
\bar Q_{s}^{(0)}\triangleq P_s^\textrm{edge}\alpha_0(H_G^2+r_s^2)^{-n_0/2},
\end{equation}
and hence
\begin{equation}\label{Psr}
P(s,r)=P_s^\textrm{edge}\bigg(\frac{H_G^2+r^2}{H_G^2+r_s^2}\bigg)^{n_0/2}.
\end{equation}
In other words, the TP of inner UEs with SF $s$ and $r<r_s$ is \textit{reduced} so that their average received power at the receiving GW equals to that of the UE at the zone edge, which achieves both fairness and power savings.

Under the above UE partitioning and power control, we derive the packet success probability in the following. 
The average interference $\bar I_s=\bar I_s^{(0)}$ is given by \eqref{barIs}, where the instantaneous interference $I_s(t)$ in \eqref{Ist} is reduced to
\begin{equation}\label{Ist0}
I_s(t)=I_s^{(0)}(t)= \sum_{k\in\mathcal{K}_s^{(0)}(t)\setminus\{0\}} \bar Q_s^{(0)} \zeta_k.
\end{equation}
Note that the UEs $k\in\mathcal{K}_s^{(0)}$ reside in the ring region $\mathcal{A}_s$, whose locations $\bold w_k$ and packet initiation time instant $t_k$ are both random, rendering the set $\mathcal{K}_s^{(0)}(t)$ difficult to model and analyze.
For simplicity, assume that the packet arriving time instants $t_k, k\in\mathcal{K}_s^{(0)}$ follow the Poisson arrival process with arrival rate $\rho_s$.
In this paper, we adopt the Poisson rain model for the UEs $k\in\mathcal{K}_s^{(0)}(t)$, which forms a space-time HPPP $\Phi_s\triangleq\{(\bold w_k, t_k), k\in\mathcal{K}_s^{(0)}\}$. We may think of $\bold w_k$ ``born" at time instant $t_k$ transmitting a packet during time interval $[t_k,t_k+T_s)$ and ``disappearing" immediately after. The HPPP $\Phi_s$ has a density $\lambda \rho_s$, which corresponds to the space-time frequency of channel access.
In the sequel, we derive the Laplace transform of the interference $\bar I_s^{(0)}$ based on the formula for the Laplace functional of the HPPP:\footnote{The formula can be extended for non-homogeneous PPP by integrating over the density measure in the considered space instead of fixed density.}
\newtheorem{lemma}{Lemma}
\begin{lemma}[Fact A.3 in \cite{NonSlottedINFOCOM2010}]\label{lem1}
Consider a generic shot-noise $J\triangleq\sum_{Y_k\in\Pi}f(\zeta_k,Y_k)$ generated by some HPPP $\Pi$ with density $\Lambda$, response function $f(\cdot,\cdot)$ and i.i.d. marks $\zeta_k$ distributed as a generic RV $\zeta$. Then the Laplace transform of $J$ is given by
\begin{equation}\label{ShotNoiseLT}
\mathcal{L}_J(z)=\exp\bigg\{-\Lambda \int\big(1-\mathbb{E}_\zeta\{e^{-zf(\zeta,y)}\}\big)\diff y\bigg\},
\end{equation}
where the integral is evaluated over the whole state space on which $\Pi$ lives.
\end{lemma}


Based on Lemma \ref{lem1}, we derive the Laplace transform of the average interference $\bar I_s^{(0)}$ in our setting with power control and bounded UE distribution field.
The results are summarized in the following proposition:
\newtheorem{prop}{Proposition}
\begin{prop}\label{prop1}
The Laplace transform of the average interference $\bar I_s^{(0)}$ from the typical cell 0 defined by \eqref{barIs} and \eqref{Ist0} under the Poisson rain model is given by
\begin{equation}\label{LaplaceIsSingle}
\mathcal{L}_{\bar I_s^{(0)}}(z)=\exp\bigg\{\frac{-2\lambda \Delta_s}{1-\Delta_s} E_s^{(0)}(z) \bigg\},
\end{equation}
where $E_s^{(0)}(z)\triangleq A_s\bigg(1+\frac{1}{\bar Q_s^{(0)} z}\ln\frac{1}{1+\bar Q_s^{(0)} z}\bigg)$.
\end{prop}

\textit{Proof:} For the average interference $\bar I_s$, it follows from \eqref{barIs} and \eqref{Ist0} that
\begin{equation}\label{barIsShot}
\bar I_s= \frac{1}{T_s}\int_{0}^{T_s} \sum_{k\in\mathcal{K}_s(t)\setminus\{0\}} \bar Q_s^{(0)} \zeta_k \diff t\stackrel{\textrm{dist.}}{=}\sum_{(\bold w_k, t_k)\in\Phi_s}\zeta_k h(t_k)\bar Q_s^{(0)},
\end{equation}
where the second equality is due to swapping of integration and summation, and the fact that the HPPP remains equal in distribution after removing one point; and 
\begin{equation}
h(t_k)\triangleq\frac{1}{T_s}\int_{0}^{T_s} \bold{1}(t_k\leq t<t_k+T_s) \diff t=\frac{\max\big(T_s-|t_k|,0\big)}{T_s},
\end{equation}
which represents the time ratio of UE $k$'s packet overlapping with the reference packet in the time interval $[0, T_s]$,
where $\bold{1}(\cdot)$ is the indicator function.

Note that \eqref{barIsShot} matches with the shot-noise definition with the response function $\zeta_k h(t_k)\bar Q_s^{(0)}$, and thus it follows from Lemma \ref{lem1} that
\begin{align}
\mathcal{L}_{\bar I_s}(z)&=e^{-\lambda\rho_s \int_{-\infty}^{\infty}\int_{0}^{2\pi}\int_{r_{s-1}}^{r_s}\big(1-\mathbb{E}_\zeta\{e^{-z\zeta h(t)\bar Q_s^{(0)}}\}\big)r\diff r\diff \phi \diff t}\notag\\
&=e^{-\lambda\rho_s A_s \int_{-\infty}^{\infty}\big(1-\mathbb{E}_\zeta\{e^{-z\zeta h(t)\bar Q_s^{(0)}}\}\big) \diff t}\notag\\
&\stackrel{(a)}{=}e^{-\lambda\rho_s A_s \int_{-\infty}^{\infty}\big(1-\frac{1}{1+zh(t)\bar Q_s^{(0)}}\big) \diff t}\notag\\
&=e^{-2\lambda\rho_s A_s T_s\big(1+\frac{1}{\bar Q_s^{(0)} z}\ln\frac{1}{1+\bar Q_s^{(0)} z}\big)},\label{LaplaceDerive}
\end{align}
where $A_s$ is given in \eqref{Area}; and $(a)$ is due to the Laplace transform of exponentially distributed $\zeta$ with mean $1$, which is given by $\mathcal{L}_{\zeta}(z')\triangleq \frac{1}{1+z'}$. Therefore, Proposition \ref{prop1} follows by substituting \eqref{rho_s} into \eqref{LaplaceDerive}.$\blacksquare$

\textit{Remark 3:} For the case with discrete power levels $P_k\in\{p_1,\cdots,p_L\}$, the continuous power level in \eqref{Psr} can be rounded to its nearest allowed power level $p_l$. As a result, the UEs with SF $s$ adopting a certain TP level $p_l$ make up a group $\mathcal{K}_{s,l}$ which reside in the distance range $(r_{s,l-1},r_{s,l}]$ from GW 0. In this case, the interference power in \eqref{Ist} is given by 
\begin{equation}\label{IstDiscrete}
I_s(t)\triangleq \sum_{k\in\mathcal{K}_s(t)\setminus\{0\}} P_k g_k=\sum_{l=1}^{L}\sum_{k\in\mathcal{K}_{s,l}(t)\setminus\{0\}} p_l \bar g_k \zeta_k.
\end{equation}
Therefore, following similarly the proof of Proposition \ref{prop1}, the interference Laplace transform in \eqref{LaplaceDerive} can be generalized as
\begin{align}
\mathcal{L}_{\bar I_s}(z)&=\prod_{l=1}^{L} e^{-2\pi\lambda\rho_s \int_{-\infty}^{\infty}\int_{r_{s,l-1}}^{r_{s,l}} \big(1-\mathbb{E}_\zeta\{e^{-z\zeta h(t) p_l\bar{g}(r)}\}\big)r  \diff r\diff t}\notag\\
&=e^{-2\pi\lambda\rho_s \sum_{l=1}^{L} \int_{r_{s,l-1}}^{r_{s,l}} \int_{-\infty}^{\infty}\big(1- \frac{1}{1+zh(t)p_l\bar{g}(r)} \big)r  \diff t\diff r}\notag\\
&=e^{-4\pi T_s\lambda\rho_s \sum_{l=1}^{L} \int_{r_{s,l-1}}^{r_{s,l}}  \big( 1+\frac{1}{p_l\bar{g}(r) z}\ln\frac{1}{1+p_l\bar{g}(r) z}  \big) r  \diff r},\label{LaplaceDeriveDiscrete}
\end{align}%
where $\bar{g}(r)\triangleq\alpha_0 (H_G^2+r^2)^{-n_0/2}$ denotes the average channel power gain of a UE $k$ at distance $r$ from GW 0.
Note that the formula in \eqref{LaplaceDeriveDiscrete} does not admit a closed-form and yet can be evaluated numerically by simple integrals.
For simplicity, continuous power control is considered in the rest of the paper in order to characterize the achievable performance upper bound, whereas the case with discrete power levels is investigated in Section \ref{SectionSimulation} to illustrate the performance loss.$\blacksquare$
\newline

For the UEs $k\in\mathcal{K}_s^{(0)}$ with i.i.d. fading $\zeta_k\stackrel{\textrm{dist.}}{=}\zeta$ and under the above power control, their SNR and SIR are both equal in distribution, respectively, and hence they have equal packet success probability, for which a closed-form lower-bound expression can be obtained based on \eqref{Psuc0}, given by
\begin{align}\label{PsusFinal}
\textrm{P}_{s,0}^{\textrm{suc}}&\geq e^{-\bar\eta_s\sigma^2/\bar Q_{s}^{(0)}}\mathcal{L}_{\bar I_s}\big(\bar\gamma_s/\bar Q_{s}^{(0)}\big)\notag\\
&=\exp\bigg\{\frac{-\sigma^2\bar\eta_{s}}{\bar Q_{s}^{(0)}}-\frac{2\lambda \Delta_s A_s C_s^{(0)}}{1-\Delta_s}\bigg\},
\end{align}
where $C_s^{(0)}\triangleq \big(1+\frac{1}{\bar\gamma_s}\ln\frac{1}{1+\bar\gamma_s}\big)$ is a constant.
Therefore, the common throughput of UEs with SF $s$ is lower bounded by
\begin{align}\label{ThetasFinal}
\bar \theta_s&\triangleq R_s\Delta_s\textrm{P}_{s,0}^{\textrm{suc}}\notag\\
&\geq R_s\Delta_s\exp\bigg\{\frac{-\sigma^2\bar\eta_{s}}{\bar Q_{s}^{(0)}}-\frac{2\lambda \Delta_s A_s C_s^{(0)}}{1-\Delta_s}\bigg\}\triangleq \bar \theta_s^\textrm{lb}.
\end{align}

Note that in the interference-limited scenario, the throughput lower-bound $\bar \theta_s^\textrm{lb}$ in \eqref{ThetasFinal} is tight, which has a tractable closed-form that reveals insights about the effects of key system parameters.
First, $\bar \theta_s^\textrm{lb}$ is monotonically increasing with the equalized received power $\bar Q_s^{(0)}$ in \eqref{Qs}, which suggests $P_s^{\textrm{edge}}=P_\textrm{max}$ to be adopted by the zone-edge UE. 
Second, $\bar \theta_s^\textrm{lb}$ is monotonically decreasing with the zone-edge distance $r_s$, which affects not only the average received power $\bar Q_s^{(0)}$ at the zone edge, but also the zone area $A_s$ (and hence the number of UEs in the zone).
Third, it can be verified that $\bar \theta_s^\textrm{lb}$ first increases and then decreases with $\Delta_s$, which facilitates efficient solutions to achieve the maximum throughput for the UEs with SF $s$. Also note that $\bar \theta_s^\textrm{lb}$ goes to 0 as $\Delta_s\rightarrow 1$, which is consistent with the practical scenario where all UEs keep transmitting all the time.
Finally, besides for system optimization, the throughput formula \eqref{ThetasFinal} is itself useful as a tractable analytical result for quick performance evaluation in LoRa networks, while its deriving method is extendable to other general settings, e.g., the multiple-GW scenario in the next section.

As a result, (P1) can be re-cast into the following problem:
\begin{align}
\mathrm{(P2)}: \underset{
\begin{subarray}{c}
  \bar\theta, \bold r\\
  \Delta_s, s\in\mathcal{S}
  \end{subarray}
}{\max}& \quad\bar\theta \notag\\
             \text{s.t.}\quad&\bar\theta_s\geq \bar\theta,\quad s\in\mathcal{S},\label{ConstraintTheta1}\\ 
             &0\leq \Delta_s\leq \Delta_\textrm{max},\quad s\in\mathcal{S},\label{ConstraintDelta1}      
\end{align}
where the optimal solution to (P2) is denoted as $(\bold r^*;\Delta_s^{*}, s\in\mathcal{S})$, and the corresponding max-min (average) throughput denoted as $\bar\theta^{*}$.
In order to solve (P2), we first obtain an estimate of the optimal duty cycle under given $\bold r$ for each SF $s$ based on the throughput lower bound in \eqref{ThetasFinal}, given by
\begin{equation}\label{DutyCycleOpt}\small
\Delta_s^*(\bold r)\approx\min\bigg\{\Delta_\textrm{max},1+\lambda A_s C_s^{(0)}-\sqrt{\lambda A_s C_s^{(0)}(2+\lambda A_s C_s^{(0)})}\bigg\},
\end{equation}
which is obtained by finding the root of the first-order derivative on $\Delta_s$ in \eqref{ThetasFinal} along with the constraint \eqref{ConstraintDelta1}.
It can be further verified that the corresponding throughput $\bar\theta_s^*(\bold r)$ is decreasing with the zone area $A_s$, which itself is decreasing with $r_{s-1}$ but increasing with $r_s$ as given in \eqref{Area}.
\rev{This corresponds to the fact that the achievable common throughput $\bar\theta_s^*$ in each SF group $s$ is decreasing as more UEs are partitioned into this group.}

\rev{Based on such monotonicity, we propose the \textit{Iterative Balancing (IB)} method to find the optimal partitioning threshold $\bold r^*$ to achieve the average max-min throughput $\bar\theta^*$.
Specifically, in each iteration, we find two neighboring zones $\mathcal{A}_s$ and $\mathcal{A}_{s+1}$ which have the largest throughput gap $\vartheta_{\textrm{max}}$, and then tune $r_s$ with others in $\bold r$ fixed, such that the throughput gap is eliminated or reduced to the best extend.
The iterations continue until $\vartheta_{\textrm{max}}$ is less than a certain threshold $\epsilon$, upon which the max-min throughput is achieved whereby the achievable common throughput $\bar\theta_s^*$ of each SF group $s$ is bounded within a difference interval of width $\epsilon$.
The optimality of the IB method can be justified based on the following property of the investigated problem, namely, the max-min throughput $\bar\theta^*$ is unique, with the achievable common throughput $\bar\theta_s^*=\bar\theta^*$ equal for all SFs $s$ used\footnote{In the case with low UE density and/or small cell radius, the highest SF(s) might not be used due to its low data rate.}.
Such a property can be proved by contradiction. First, assume otherwise that $\bar\theta_s^*>\bar\theta_{s+1}^*$ for a certain SF $s$. Then we can tune up $r_s$ such that more UEs are partitioned from SF $s+1$ into SF $s$, and as a result the lower throughput $\bar\theta_{s+1}^*$ can be increased. Therefore, by induction, the higher throughput in a certain SF could help increase the lower throughput in other SFs, which leads to increased minimum throughput and thus poses contradiction.
Second, assume that the current max-min throughput $\bar\theta^*$ is achieved by the partitioning distance $\bold r^*$, and there exists a higher one $\bar\theta^{*'}>\bar\theta^*$. Then for the lowest SF 7, to achieve $\bar\theta_7=\bar\theta^{*'}>\bar\theta^*$, the partitioning distance $r_7$ needs to be reduced. Similarly for other higher SFs, the corresponding partitioning distance $r_s$ needs to be reduced in order to achieve $\bar\theta^{*'}$. By induction, this would lead to a contradicting result that the served cell radius is smaller than $r_c$.
Therefore, the property mentioned above is proved.

Similarly, the convergence of the IB method can be justified by induction. For all the SFs used, sort their currently achieved common throughput $\bar\theta_s$ by descending order, and define $\vartheta_1\triangleq \max\limits_s\bar\theta_s -\min\limits_s \bar\theta_s$, i.e., the gap between the highest and lowest throughputs achieved so far.
In each iteration of the IB method, if any of the SF $s=\arg\max \bar\theta_s$ or $s=\arg\min \bar\theta_s$ is encountered, then the gap $\vartheta_1$ would be reduced since either the highest throughput is reduced or the lowest throughput is increased, or both. Otherwise not encountered, we can further define $\vartheta_2$ as the gap between the second highest throughput and the second lowest throughput. Similarly, the gap $\vartheta_2$ is either reducing or the induction continues, by the end of which only one gap exists and will be eliminated, thus returning to the previous level of induction and reducing the corresponding gap $\vartheta_i$. As a result, the bounding gap $\vartheta_1$ is ultimately reduced which leads to convergence of the IB method.}

\begin{algorithm}[H]\caption{Computing the partitioning distance threshold $\bold r$ for SF allocation by IB method}\label{AlgIB}
\begin{small}
\rev{\begin{algorithmic}[1]
\STATE Initialize $\bold r$, where $r_s\leq r_{s+1}\leq r_c, s\in\mathcal{S}$.
\REPEAT
\STATE For each SF $s\in\mathcal{S}$, obtain the throughput $\bar\theta_s$ under a given duty cycle $\Delta_s$ ($\Delta_s$ could be initialized as the estimated optimal duty cycle in \eqref{DutyCycleOpt}, and fine-tuned for the optimal).
\STATE Find the largest throughput gap $\vartheta_{\textrm{max}}=\max\limits_{7\leq s\leq 11}|\bar\theta_{s}-\bar\theta_{s+1}|$ and corresponding index $s_0$.
\STATE Adjust $r_{s_0}$ while keeping others in $\bold r$ fixed, subject to $r_{s_0-1}\leq r_{s_0}\leq r_{s_0+1}$ and $r_{s_0}\leq r_{s_0,\textrm{max}}$:

\IF{$\bar\theta_{s_0}\leq\bar\theta_{s_0+1}$}
\STATE Tune $r_{s_0}$ down such that $\bar\theta_{s_0}=\bar\theta_{s_0+1}$; 
\ELSE
\STATE Tune $r_{s_0}$ up such that $\bar\theta_{s_0}=\bar\theta_{s_0+1}$, or until $r_{s_0}=\min\{r_{s_0+1},r_{s_0,\textrm{max}}\}$.
\ENDIF

\UNTIL{$\vartheta_{\textrm{max}}<\epsilon$, or none of the throughput gaps can be reduced, or the maximum number of iterations $N_\textrm{max}$ is reached.}
\end{algorithmic}}
\end{small}
\end{algorithm}



\rev{The IB method is summarized in Algorithm \ref{AlgIB}.
Note that we place a bound $N_\textrm{max}$ on the maximum number of iterations in order to limit the worse-case running time.
Also, as an option, we can further define an upper limit $r_{s,\textrm{max}}$ for $r_s$ in practice, which could be set, for example, as the maximum range reachable by SF $s$ under path-loss only. 
For the tuning process in Line 7 or 9, the worst-case complexity is given by $O\big(\log_2(1/\epsilon)\big)$ if bisection search is used, which runs fast since the underlying computation is based on our derived (semi-) closed-form expressions.
As a result, the overall complexity of the IB method is given by $O\big(N_\textrm{max}\log_2(1/\epsilon)\big)$, although the algorithm runs much faster in practice than such worse-case time complexity.}\footnote{\rev{In our simulations, a value of $N_\textrm{max}=50$ would suffice, and such computation typically completes within a few minutes on a laptop computer even for the multi-cell scenarios in Sections \ref{SectionMultiGW} and \ref{SectionMultiGWreception}.}}
Finally, note that for the case where the duty cycle is pre-determined and cannot be adjusted, the proposed IB method can still be applied to achieve the corresponding average max-min throughput given the duty cycle.


\section{Frequency Reuse Schemes in Multi-GW LoRa Networks}\label{SectionMultiGW}

\begin{figure}
        \centering
        \hspace{-10pt}
        \begin{subfigure}[b]{0.55\linewidth}
                \includegraphics[width=1\linewidth,  trim=0 0 0 0,clip]{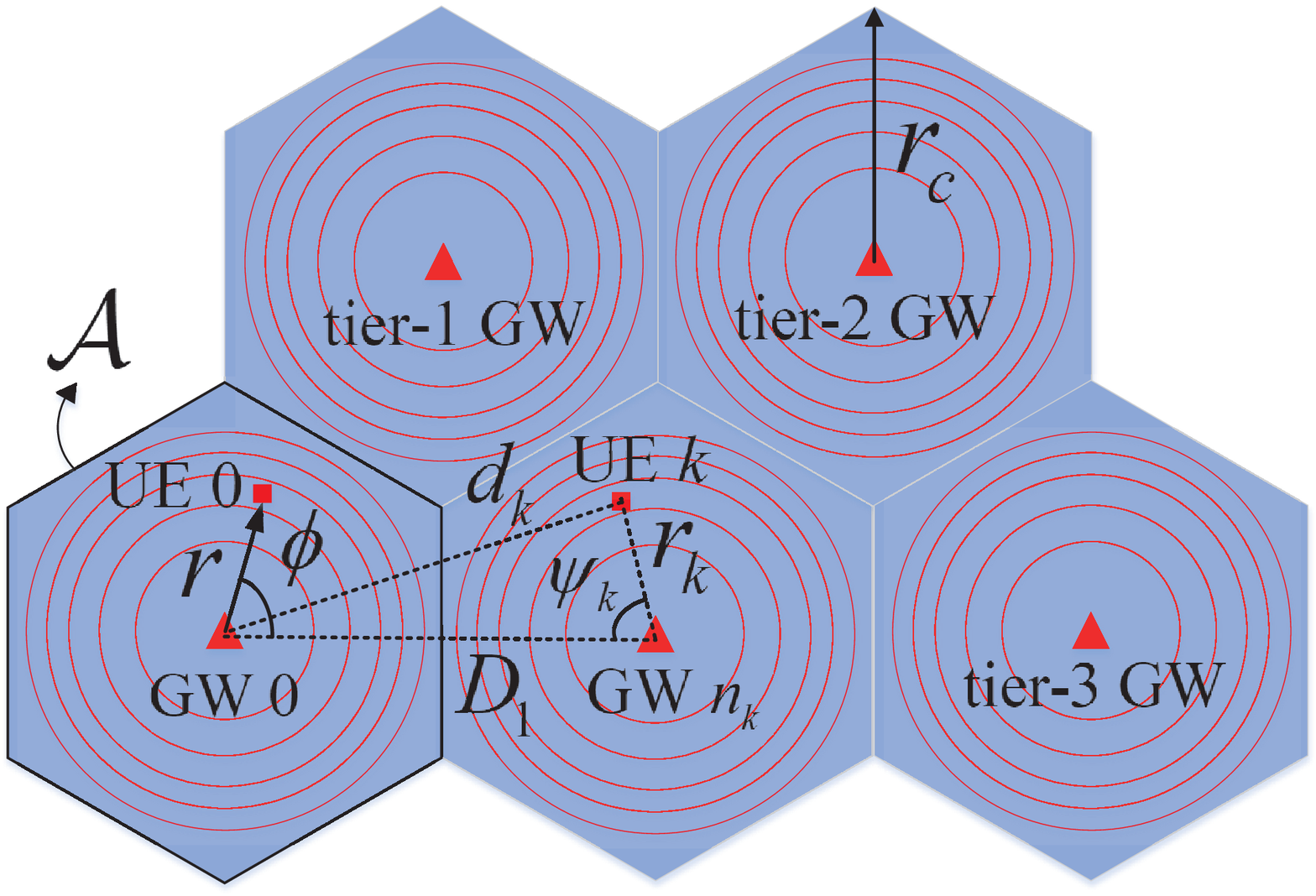}
\caption{\vspace{0ex}}\label{visio_1_Reuse}
        \end{subfigure}%
        \begin{subfigure}[b]{0.45\linewidth}
                \includegraphics[width=1\linewidth,  trim=0 0 0 0,clip]{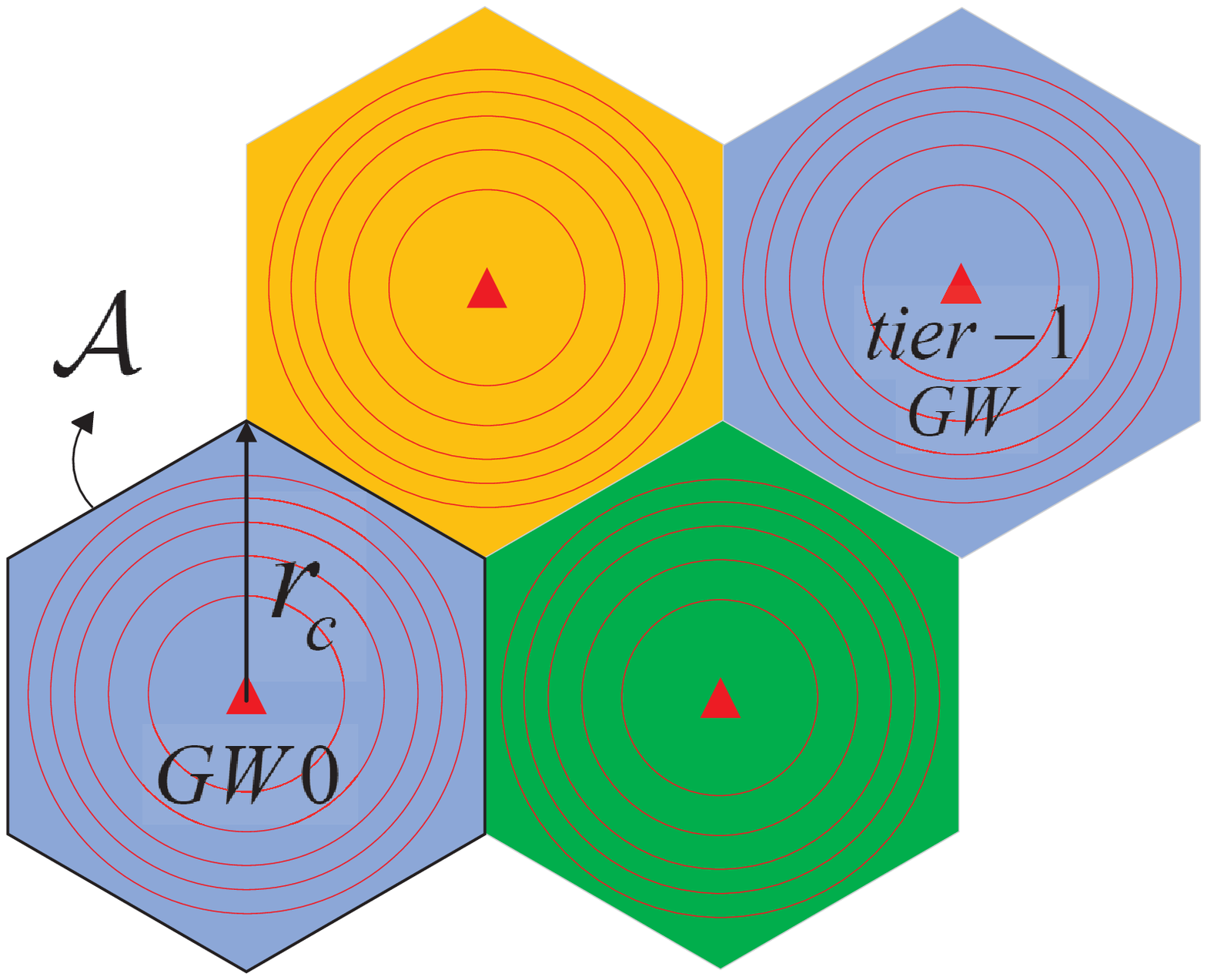}
\caption{\vspace{0ex}}\label{visio_1_3Reuse}
        \end{subfigure}%
        \caption{(a) 1-reuse. (b) 1/3 reuse.\vspace{-2ex}}\label{Reuse}
\end{figure}

In this section, we consider the multi-GW scenario in Fig. \ref{LoRa}(b), and investigate how to achieve the max-min throughput of the multi-GW LoRa network under different frequency reuse schemes and different GW density (corresponding to different cell radius $r_c$).
Denote $\lambda_\textrm{active}$ as the total UE density to be served in the considered multi-cell region during the considered time period.
To incorporate the additional interference from other cells, we consider a typical GW 0 and its cell region $\mathcal{A}$ in Fig. \ref{Reuse}, and $M$ tiers of other co-channel GWs, denoted by the sets $\mathcal{M}_m, m=1,2,\cdots,M$, where the GWs in tier $m$ have the same distance $D_m$ from GW 0.
For example, we have $D_1=\sqrt{3}r_c$, $D_2=3 r_c$ and $D_3=2\sqrt{3} r_c$ for the 1-reuse scheme in Fig. \ref{Reuse}(a).
By default, denote $m=0$ as the 0-th tier which consists of GW 0 only.

\subsection{1-Reuse}\label{Section1reuse}

Consider the full-reuse scheme ($\omega=1$) where all GWs use the full set of $\mathrm{F}$ channels, and the UE density on each channel is given by $\lambda=\lambda_\textrm{active}/\mathrm{F}$.
As a result, the overall received interference at GW 0 comes from different tiers of cells, and the expression in \eqref{Ist} is now given by
\begin{equation}\label{sumTiers}
I_s(t)\triangleq I_s^{(0)}(t)+\sum_{m=1}^{M} I_s^{(m)}(t),
\end{equation}
where $I_s^{(0)}(t)$ follows directly from \eqref{Ist0}, and $I_s^{(m)}(t)$ is the total interference from the tier-$m$ cells, which is given by
\begin{equation}
I_s^{(m)}(t)\triangleq \sum_{k\in\mathcal{K}_s^{(m)}(t)} P_k g_k=\sum_{k\in\mathcal{K}_s^{(m)}(t)} P_k \bar g_k \zeta_k,
\end{equation}
where $\mathcal{K}_s^{(m)}(t)$ denotes the set of UEs in tier-$m$ cells which are transmitting at time $t$ with SF $s$ on the typical channel.

Assume that all cells adopt the same policies of SF/TP/duty cycle control, which follow similarly from the single-cell case in Section \ref{SectionSingle}.
Denote $n_k$ as the GW in UE $k$'s residing cell.
Denote $r_k$ and $\psi_k$ as the UE $k$-GW $n_k$ distance and UE $k$-GW $n_k$-GW 0 angle, respectively, as shown in Fig. \ref{Reuse}(a).
As a result, the TP $P_k$ of UE $k$ with SF $s$ follows the power control policy in \eqref{Psr}, with $r=r_k$.
On the other hand, the average channel power $\bar g_k$ from an interfering UE $k$ in a tier-$m$ cell ($m=1,\cdots,M$) to GW 0 is given by $\bar g_k= \alpha_0 (H_G^2+d_k^2)^{-n_0/2}$, where $d_k^2$ can be expressed using the cosine law as
\begin{equation}
d_k^2=D_m^2+r_k^2-2r_k D_m\cos\psi_k.
\end{equation}

Therefore, the average interference power received at GW 0 caused by UE $k$ with SF $s$ in tier-$m$ cell is given by
\begin{align}\label{Qsrphi}
&Q_{s,k}^{(m)}\triangleq P_k \bar g_k =P_s^\textrm{edge}\bigg(\frac{H_G^2+r_k^2}{H_G^2+r_s^2}\bigg)^{\frac{n_0}{2}}\cdot\frac{\alpha_0}{(H_G^2+d_k^2)^{\frac{n_0}{2}}}\notag\\
&=P_s^\textrm{edge} \alpha_0\bigg(\frac{H_G^2+r_k^2}{(H_G^2+r_s^2)(H_G^2+D_m^2+r_k^2-2r_kD_m\cos\psi_k)}\bigg)^{\frac{n_0}{2}}\notag\\
&\triangleq Q_{s}^{(m)}(r_k,\psi_k).
\end{align}
As a result, the total interference at time $t$ from tier-$m$ cells is given by
\begin{equation}
I_s^{(m)}(t)\triangleq \sum_{k\in\mathcal{K}_s^{(m)}(t)} Q_{s,k}^{(m)} \zeta_k,
\end{equation}
which has a similar form with $I_s^{(0)}(t)$ in \eqref{Ist0}.

Based on \eqref{sumTiers},
the aggregate interference power at GW 0 averaged over one packet duration $T_s$ is then given by
\begin{equation}
\bar I_s\triangleq \frac{1}{T_s}\int_{0}^{T_s} I_s(t) \diff t=\bar I_s^{(0)}+\sum_{m=1}^{M} \bar I_s^{(m)},
\end{equation}
where $\bar I_s^{(0)}$ denotes the average total interference from cell 0, and $\bar I_s^{(m)}$ denotes the average total interference from tier-$m$ cells, $m=1,\cdots,M$, which are mutually independent.
As a result, the Laplace transform of $\bar I_s$ is given by
\begin{equation}\label{LaplaceIsAll}
\mathcal{L}_{\bar I_s}(z)\triangleq \mathbb{E}_{\bar I_s}\big\{e^{-z\bar I_s}\big\}=\mathcal{L}_{\bar I_s^{(0)}}(z)\prod_{m=1}^M \mathcal{L}_{\bar I_s^{(m)}}(z),
\end{equation}
where $\mathcal{L}_{\bar I_s^{(m)}}(z)$ denotes the Laplace transform of $\bar I_s^{(m)}$, $m=0,1,\cdots,M$.

For the typical cell 0, $\mathcal{L}_{\bar I_s^{(0)}}(z)$ follows directly from \eqref{LaplaceIsSingle} with the UE density $\lambda=\lambda_\textrm{active}/\mathrm{F}$, yet noting that the area $A_s$ may not be a ring region as in \eqref{Area}, but instead may be a slightly more complicated shape (e.g., UEs with SF 12 may reside in a region with hexagonal boundary).
For other cells in tier-$m$, we have the following proposition.
\begin{prop}\label{prop2}
	The Laplace transform of $\bar I_s^{(m)}$, $m=1,\cdots,M$ is given by
\begin{equation}\label{LaplaceDeriveOtherShort}
\mathcal{L}_{\bar I_s^{(m)}}(z)=\exp\bigg(-\frac{2\lambda \Delta_s |\mathcal{M}_m| E_s^{(m)}(z) }{1-\Delta_s}\bigg),
\end{equation}
where $|\mathcal{M}_m|$ is the number of GWs in tier $m$,\footnote{For example, we have $|\mathcal{M}_1|=|\mathcal{M}_2|=|\mathcal{M}_3|=6$ in the 1-reuse scheme in Fig. \ref{Reuse}(a).} and $E_s^{(m)}(z)$ is given by
\begin{equation}\footnotesize\label{Esmz}
E_s^{(m)}(z)\triangleq \int_{0}^{2\pi}\int_{r_{s-1}}^{r_s} \bigg(1+\frac{1}{z Q_{s}^{(m)}(r,\psi)}\ln\frac{1}{1+z Q_{s}^{(m)}(r,\psi)}\bigg)   r \diff r\diff \psi,
\end{equation}
with $Q_{s}^{(m)}(r,\psi)$ given by \eqref{Qsrphi}.
\end{prop}

\textit{Proof:} The average total interference from tier-$m$ cells is given by
\begin{equation}\small
\bar I_s^{(m)}\triangleq \frac{1}{T_s}\int_{0}^{T_s} I_s^{(m)}(t) \diff t=\frac{1}{T_s}\int_{0}^{T_s} \sum_{k\in\mathcal{K}_s^{(m)}(t)} Q_{s,k}^{(m)} \zeta_k \diff t,
\end{equation}
which comprises $|\mathcal{M}_m|$ parts of i.i.d. interference $I_{s,n}^{(m)}$ from cells $n\in \mathcal{M}_m$.
Therefore, based on similar derivations for Proposition \ref{prop1}, the Laplace transform of $\bar I_s^{(m)}$ is given by
\begin{align}
&\mathcal{L}_{\bar I_s^{(m)}}(z)=\prod_{n\in \mathcal{M}_m} \mathcal{L}_{\bar I_{s,n}^{(m)}}(z)\notag\\
&=e^{-\lambda\rho_s |\mathcal{M}_m| \int_{-\infty}^{\infty}\int_{0}^{2\pi}\int_{r_{s-1}}^{r_s}\big(1-\mathbb{E}_\zeta\big\{e^{-z\zeta h(t) Q_{s}^{(m)}(r,\psi)}\big\}\big)r\diff r\diff \psi \diff t}\notag\\
&\stackrel{(a)}{=}e^{-\lambda\rho_s |\mathcal{M}_m| \int_{0}^{2\pi}\int_{r_{s-1}}^{r_s}\int_{-\infty}^{\infty}\big(1-\frac{1}{1+z h(t) Q_{s}^{(m)}(r,\psi)}\big)r \diff t\diff r\diff \psi}\notag\\
&\stackrel{(b)}{=}e^{-\lambda\rho_s |\mathcal{M}_m|2 T_s\int_{0}^{2\pi}\int_{r_{s-1}}^{r_s}  \big(1+\frac{1}{z Q_{s}^{(m)}(r,\psi)}\ln\frac{1}{1+z Q_{s}^{(m)}(r,\psi)}\big)   r \diff r\diff \psi}\notag\\
&=e^{-2\lambda |\mathcal{M}_m| \rho_s T_s E_s^{(m)}(z)},\label{LaplaceDeriveOther}
\end{align}
where $Q_{s}^{(m)}(r,\psi)$ is given by \eqref{Qsrphi}; $(a)$ is due to the Laplace transform of exponentially distributed $\zeta$ with mean $1$; $(b)$ is by integrating over $t$, similarly to that in \eqref{LaplaceDerive};
and $E_s^{(m)}(z)$ is given by \eqref{Esmz}.
The final expression in \eqref{LaplaceDeriveOtherShort} is then obtained by substituting $\rho_s$ in \eqref{rho_s} into \eqref{LaplaceDeriveOther}.$\blacksquare$

As a result, based on \eqref{LaplaceIsSingle}, \eqref{LaplaceIsAll} and \eqref{LaplaceDeriveOtherShort}, the Laplace transform of $\bar I_s$ is then given in a semi-closed form expression:
\begin{equation}\label{LaplaceIsAllExpand}
\mathcal{L}_{\bar I_s}(z)=\exp\bigg\{\frac{-2\lambda \Delta_s}{1-\Delta_s}G_s(z) \bigg\},
\end{equation}
where $G_s(z)\triangleq E_s^{(0)}(z)+ \sum_{m=1}^M|\mathcal{M}_m| E_s^{(m)}(z)$.
Therefore, the packet success probability can be obtained based \eqref{Psuc0} with the interference Laplace transform substituted by the expression in \eqref{LaplaceIsAllExpand}, which is given by
\begin{align}\label{PsusMultiGW}
&\textrm{P}_{s,0}^{\textrm{suc}}\geq\exp\bigg\{-\frac{\bar\eta_s\sigma^2}{\bar Q_{s}^{(0)}}\bigg\}\mathcal{L}_{\bar I_s}(z)\big|_{z=\bar\gamma_s/\bar Q_{s}^{(0)}}\notag\\
&=\exp\bigg\{-\frac{\bar\eta_s\sigma^2}{\bar Q_{s}^{(0)}}-\frac{2\lambda \Delta_s}{1-\Delta_s} G_s\bigg(\frac{\bar\gamma_s}{\bar Q_{s}^{(0)}}\bigg)\bigg\}.
\end{align}

Based on \eqref{PsusMultiGW}, an estimate of the optimal duty cycle to maximize the common throughput $\bar \theta_s=R_s\Delta_s\textrm{P}_{s,0}^{\textrm{suc}}$ can be derived similarly as in \eqref{DutyCycleOpt}, which is given
\begin{equation}\small\label{DutyCycleOptMultiGW}
\Delta_s^*(\bold r)\approx\min\big\{\Delta_\textrm{max},1+\lambda G_s(z_0)-\sqrt{\lambda G_s(z_0)[2+\lambda G_s(z_0)]}\big\},
\end{equation}
with $z_0\triangleq \bar\gamma_s/\bar Q_{s}^{(0)}$.
As a result, the derived formulas in \eqref{DutyCycleOpt} and \eqref{DutyCycleOptMultiGW} can be applied to approximate the optimal duty cycle within each SF group, which provides quick reference for traffic shaping under given UE/GW densities.

\rev{Finally, thanks to similar monotonic property of the achievable common throughput $\theta_s^*$ as discussed in Section \ref{SectionSingle}, we can tune the partitioning distance threshold $\bold r$ for SF allocation 
to achieve the average max-min throughput $\bar\theta^*$, based on the IB method in Algorithm \ref{AlgIB}.
}

\subsection{$1/\mathrm{F}$-Reuse}\label{Sectoin13}
Consider the $1/\mathrm{F}$-reuse scheme where each cell uses only one of the $\mathrm{F}$ channels based on a certain pattern. As a result, on one typical channel, the UE density in the co-channel cells is given by $\lambda=\lambda_\textrm{active}$, which is $\mathrm{F}$ times that of the 1-reuse scheme and hence more interference arises from a co-channel cell at a given distance away. 
On the other hand, however, the distance $D_m$ of tier-$m$ GWs also becomes larger thanks to the frequence reuse pattern, hence resulting in weaker interference generated per UE.
Therefore, there is a general trade-off between the density and distance of interfering UEs in different frequency reuse patterns.
For the case with $\mathrm{F}=3$ channels, an illustrative example is shown in Fig. \ref{Reuse}(b), with $D_1=3r_c$ for tier-1 co-channel GWs in the 1/3-reuse scheme.

Despite the density and distance of interefering UEs, the analytical framework follows directly from that in Section \ref{Section1reuse}, including formulas for the interference, packet success probability and throughput, as well as the solution method. Note that the reuse patterns with different $\mathrm{F}$ or other reuse factors can be similarly analyzed by our analytical framework.

\subsection{LoRa-FFR}

\begin{figure}
        \centering
        \hspace{-10pt}
        \begin{subfigure}[b]{0.475\linewidth}
                \includegraphics[width=1\linewidth,  trim=0 0 0 0,clip]{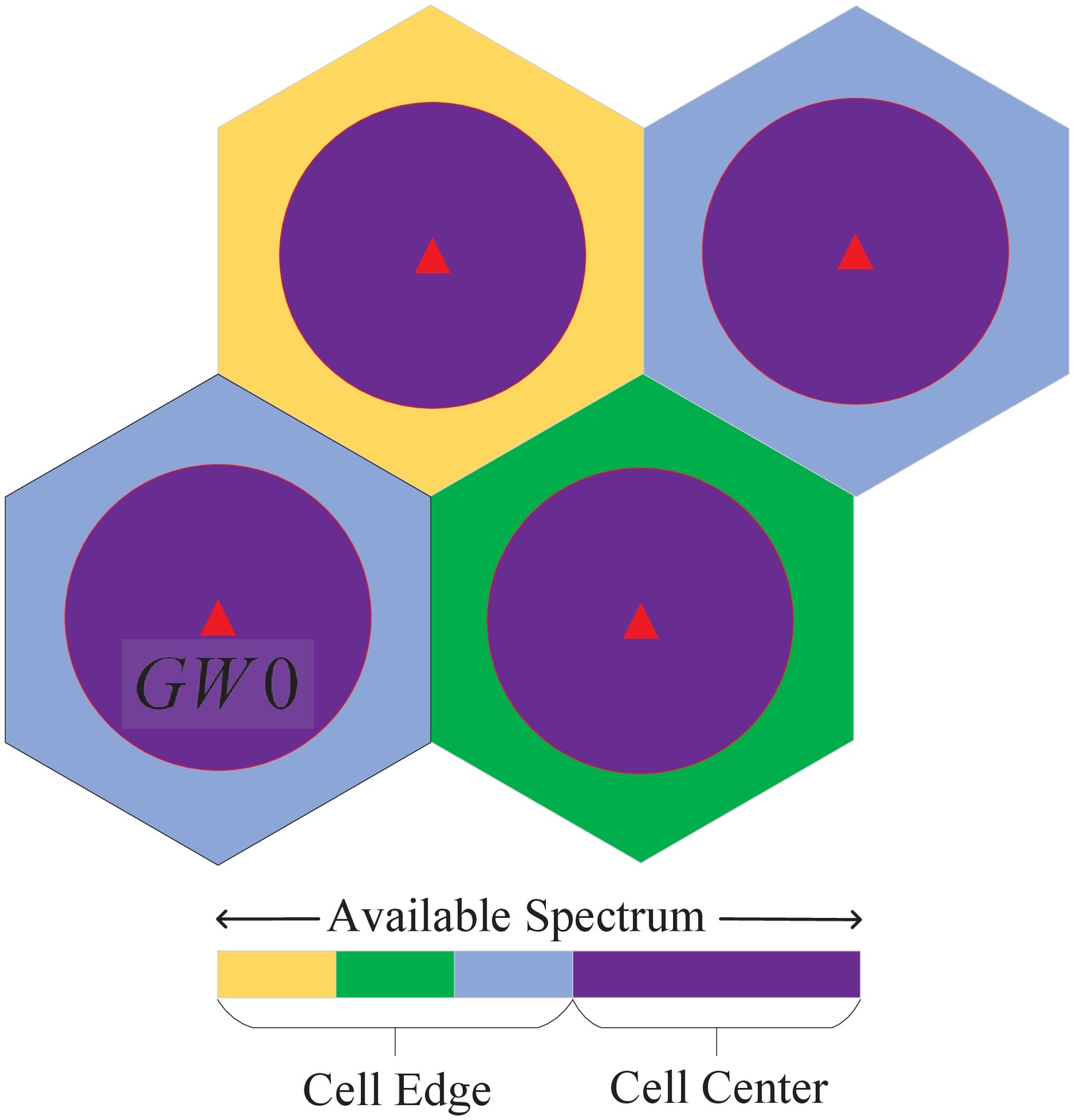}
\caption{\vspace{0ex}}\label{visio_Classic}
        \end{subfigure}%
        \begin{subfigure}[b]{0.525\linewidth}
                \includegraphics[width=1\linewidth,  trim=0 0 0 0,clip]{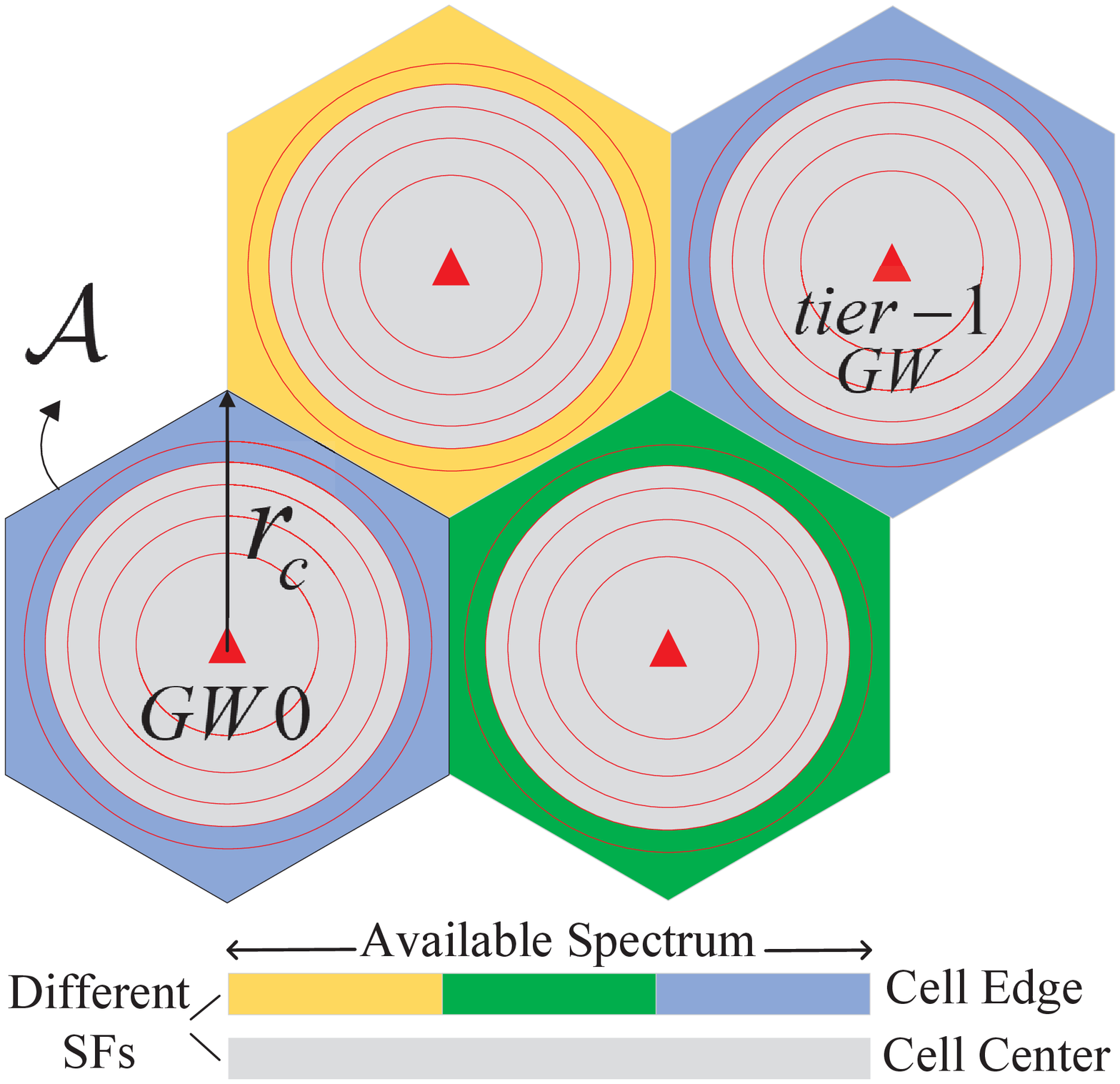}
\caption{\vspace{0ex}}\label{visio_fracrionReuse}
        \end{subfigure}%
        \caption{(a) Classic FFR. (b) LoRa-FFR.\vspace{-2ex}}\label{FReuse}
\end{figure}

Fractional frequency reuse (FFR) is a classic frequency reuse scheme which divides each cell into cell-center and cell-edge zones which are allocated with orthogonal bands. Moreover, the cell-edge zones in neighboring cells are also allocated with orthogonal bands in order to mitigate the inter-cell interference, since the cell-edge UEs are closer to other cells. An example of the classic FFR scheme is shown in Fig. \ref{FReuse}(a). 

In this subsection, we propose a modified FFR scheme for LoRa networks, termed as \textit{LoRa-FFR}, which also allocates orthogonal bands to different cell-edge zones in neighboring cells, as shown in Fig. \ref{FReuse}(b).
In contrast, however, there is an additional degree of design freedom, i.e., the pseudo-orthogonal SFs, which naturally separate the cell-center and cell-edge zones according to the distance-based SF allocation policy as in Fig. \ref{LoRa}(a) and Fig. \ref{Reuse}.
As a result, the cell-center zones can now reuse the full spectrum band, thus further improving the spectrum efficiency.
In this case, on one typical channel, the UE density in the cell-edge zones is still given by $\lambda=\lambda_\textrm{active}$, while it is reduced to $\lambda=\lambda_\textrm{active}/\mathrm{F}$ in the cell-center zones thanks to the pseudo-orthogonal SFs in LoRa networks.
Similarly, the analysis for the cell-center and cell-edge zones follows directly from that in Sections \ref{Section1reuse} and \ref{Sectoin13}, respectively.

\section{Macro Diversity of Multi-GW Reception}\label{SectionMultiGWreception}

In Section \ref{SectionMultiGW}, we have characterized the packet success probability which considers uplink packet reception at GW 0 only.
Different from the conventional cellular network, a LoRa uplink packet can be potentially received by multiple nearby GWs. The uplink transmission is successful if the packet can be decoded by one of the GWs.
Thanks to such macro diversity, each uplink packet gets additional chances to be received by nearby GWs in other cells, especially for cell-edge UEs which are closer to other GWs.
Note that we have proposed the complete channel inversion power control for the single-GW reception case in the above sections, which equalizes the received signal power at GW 0 (and hence packet success probability) for the UEs with the same SF $s$.
In the multi-GW reception scenario, the UEs at the outer-boundary of the ring region (i.e., at distance $r_s$ from GW 0) tend to have a higher packet success probability than those at the inner-boundary due to the macro diversity.
Therefore, to maximize the minimum throughput in each ring region, we propose the \textit{distance-proportional fractional power control} by introducing a power control factor $\beta\in[0,1]$ in \eqref{Psr}, i.e.,
\begin{equation}\label{PsrFractional}
P(s,r)=P_s^\textrm{edge}\bigg(\frac{H_G^2+r^2}{H_G^2+r_s^2}\bigg)^{\frac{n_0}{2} \beta},
\end{equation}
whereby $\beta=0$ corresponds to the special case with constant power $P(s,r)=P_s^\textrm{edge}$, while $\beta=1$ reduces to the case in \eqref{Psr} with complete channel inversion.
As a result, the TP of the typical UE 0 at location $\bold w_0\triangleq (r,\phi)$ with SF $s$ is then given by $P_0=P(s,r)$, where $(r,\phi)$ denotes the polar coordinate centered at GW 0, as shown in Fig. \ref{Reuse}(a).

For the channel between UE 0 and GW $n$, the channel power gain can be modelled as
$g_{0,n}=\bar g_{0,n}\zeta_{0,n}$, where $\zeta_{0,n}\sim \textrm{Exp}(1)$ and $\bar g_{0,n}\triangleq\alpha_0 (H_G^2+d_{0,n}^2)^{-n_0/2}$ is the average channel power gain, with $d_{0,n}\triangleq \|\bold w_0-\bold v_n\|$ denoting the corresponding horizontal distance, and $\bold v_n$ denoting the horizontal location of GW $n$. 
Based on the cosine law, the square of the distance $d_{0,n}$ can be expressed in the polar coordinate system as
\begin{equation}
d_{0,n}^2=r^2+\|\bold v_n\|^2-2r\|\bold v_n\|\cos(\phi-\angle \bold v_n).
\end{equation}
Therefore, the average channel power gain $\bar g_{0,n}$ is also a function of UE 0's location $\bold w_0\triangleq (r,\phi)$.

Denote $\eta_{s,0,n}$ and $\gamma_{s,0,n}$ as the SNR and SIR of a reference packet sent by the typical UE 0 with SF $s$ received at GW $n$.
Following the definition of packet success probability at GW 0 in \eqref{PsusTrue}, we can similarly obtain the packet success probability at GW $n$, which is given by
\begin{align}\label{Psucn}
&\textrm{P}_{s,0,n}^{\textrm{suc}}(\bold w_0)\triangleq\mathbb{P}\big\{\{\eta_{s,0,n}\geq\bar\eta_s\} \&\{\gamma_{s,0,n}\geq\bar\gamma_s\}\big\}\notag\\
&\geq e^{-\bar\eta_s\sigma^2/(P_0\bar g_{0,n})}\mathbb{E}_{\bar I_{s,n}}\big\{e^{-\bar\gamma_s\bar I_{s,n}/(P_0\bar g_{0,n})}\big\},
\end{align}
where $\bar I_{s,n}$ denotes the aggregate interference power at GW $n$ averaged over one packet duration $T_s$, similar to the definition of $\bar I_{s}$ for GW 0 in \eqref{barIs}.
The term $e^{-\bar\eta_s\sigma^2/(P_0\bar g_{0,n})}$ represents the packet success probability based on the SNR condition, while the term $\mathbb{E}_{\bar I_{s,n}}\big\{e^{-\bar\gamma_s\bar I_{s,n}/(P_0\bar g_{0,n})}\big\}\triangleq \mathcal{L}_{\bar I_{s,n}}\big(\bar\gamma_s/(P_0\bar g_{0,n})\big)$ represents that based on the SIR condition. In order to obtain $\textrm{P}_{s,0,n}^{\textrm{suc}}$, we need to derive the interference Laplace transform $\mathcal{L}_{\bar I_{s,n}}(\cdot)$ first.


For GW 0, we have assumed that its received interference comes from $M$ tiers of other GWs centered at GW 0.
Similarly, by symmetry, we assume that the interference received by GW $n$ comes from its own $M$ tiers of surrounding GWs.
As a result, the interference $\bar I_{s,n}$ received by GW $n$ has the same statistical distribution as that of the interference $\bar I_{s}$ received by GW 0.
Therefore, the Laplace transform of $\bar I_{s,n}$ can be obtained similarly to that in \eqref{LaplaceIsAllExpand}, with only slight modification of $Q_{s}^{(m)}(r,\psi)$ in \eqref{Qsrphi} to account for the fraction power control in \eqref{PsrFractional}, i.e.,
\begin{equation}\label{QsrphiFractional}
Q_{s}^{(m)}(r,\psi) =\frac{\alpha_0 P_s^\textrm{edge}\bigg(\frac{H_G^2+r^2}{H_G^2+r_s^2}\bigg)^{\frac{n_0}{2} \beta}}{(H_G^2+D_m^2+r^2-2r D_m\cos\psi)^{\frac{n_0}{2}}}.
\end{equation}
As a result, the packet success probability at GW $n$ can then be obtained by substituting the interference Laplace transform $\mathcal{L}_{\bar I_{s,n}}(\cdot)$ into \eqref{Psucn}.


Finally, with multi-GW reception, a LoRa uplink packet fails if and only if none of the GWs can decode the packet.
For analytical tractability, we assume that the interferences received by different GWs are mutually independent, as is justified in \cite{LoRaMacroDiversity}.
As a result, given the location $\bold w_0$ of the typical UE 0, its overall packet success probability with SF $s$ is given by
\begin{equation}
\textrm{P}_{s,0}^{\textrm{suc}}(\bold w_0)\triangleq 1-\prod_{n\in\mathcal{M}} \big(1-\textrm{P}_{s,0,n}^{\textrm{suc}}(\bold w_0)\big).
\end{equation}
Therefore, the average packet success probability for the UEs with SF $s$ can be obtained by integrating over the distribution of $\bold w_0$, i.e.,
\begin{equation}\label{PsusInt}
\bar{\textrm{P}}_{s,0}^{\textrm{suc}}\triangleq \frac{1}{A_s} \int_{\phi=0}^{2\pi}\int_{r_{s-1}}^{r_{s}} \textrm{P}_{s,0}^{\textrm{suc}}(\bold w_0) r\diff r\diff \phi.
\end{equation}
The average throughput of the typical UE 0 with SF $s$ under multi-GW reception is thus given by
\begin{equation}\label{theta0n}
\bar\theta_{s,0}\triangleq R_s\Delta_s\bar{\textrm{P}}_{s,0}^{\textrm{suc}}.
\end{equation}
\rev{Finally, we can similarly tune the partitioning distance threshold $\bold r$ for SF allocation 
	to achieve the average max-min throughput $\bar\theta^*$, based on the IB method in Algorithm \ref{AlgIB}.
}

In summary, we have extended our analytical results of the single-cell LoRa network in Section \ref{SectionSingle} to the multi-cell LoRa network under single-GW/multi-GW reception in Sections \ref{SectionMultiGW} and \ref{SectionMultiGWreception}, respectively.
Based on these analytical results, we are able to perform quick evaluation of the throughput fairness and scalability in a large-scale LoRa network under different UE/GW densities, which are cross-validated with Monte-Carlo (MC) simulations in the next section.

\section{Numerical Results}\label{SectionSimulation}



In this section, we first perform MC simulations and verify the accuracy of our proposed analytical formulas for the average throughput in \eqref{theta0} for both the single-GW and multi-GW scenarios.
In the MC simulations, 
we first generate the location database for the set $\mathcal{K}_{\textrm{all}}$ of all UEs (including both active and inactive) in the considered area, which is a random realization of an HPPP with larger density $\lambda_{\textrm{all}}$ (e.g., $\lambda_{\textrm{all}}=10\lambda_\textrm{active}$).
Then the set $\mathcal{K}$ of active UEs per channel in the considered time period is randomly and independently drawn from $\mathcal{K}_{\textrm{all}}$.
The packet success probability is obtained by averaging over $N$ (e.g., $N=10^6$) realizations of $\mathcal{K}$, where for each realization we simulate the random packet generation and overlapping, and verify both the SNR and SIR conditions for the reference packet according to \eqref{PsusTrue}.
The following parameters are used if not mentioned otherwise: $H_G=25$ m, $\lambda=350$/km$^2$, $B=125$ kHz, $C=4/5$, $L_s=25$ Bytes, $P_{\textrm{max}}=14$ dBm, $\beta=0.9$, $n_0=3.5$, $f_c=868$ MHz, $c=3\times 10^8$ m/s, $\sigma^2=-117$ dBm, $\Delta_{\textrm{max}}=1\%$, $\epsilon=0.02$ bps, $\bar\gamma_s=6$ dB for $s\in\mathcal{S}$, and $\bar\eta_s$ is given in Table \ref{table1}.
\begin{table}[t]\scriptsize
	\caption{LoRa parameters}
	\addtolength{\tabcolsep}{-4pt}
	\renewcommand{\arraystretch}{1.1}
	\centering
	\begin{tabular}{|c|c|c|c|}
		\hline
		SF&Bit rate $R_s$(bps)&SNR threshold $\bar\eta_s$ (dB)&Max. range (m) under pathloss only\\ \hline
		7 & 5469 & -6 & 1053 \\ \hline
		8 & 3125 & -9 & 1283 \\ \hline
		9 & 1758 & -12 & 1563 \\ \hline
		10 & 977 & -15 & 1904 \\ \hline
		11 & 537 & -17.5 & 2244 \\ \hline
		12 & 293 & -20 & 2645 \\ \hline
	\end{tabular}
	\centering
	\label{table1}
	\vspace{-0.5em}
\end{table}

Based on the verified analytical formulas, we can then apply our proposed optimization scheme to achieve the average max-min throughput in the LoRa network.
Besides the average minimum (common) throughput $\bar\theta_\textrm{min}$, we define the \textit{spatial throughput} of all UEs $k\in\mathcal{K}$ in bps/m$^2$ as the ratio of the average total throughput versus the considered population area $S$, i.e.,
\begin{equation}\label{SpatialThroughput}
\Theta\triangleq \frac{1}{S}\mathbb{E}\bigg\{\sum_{s\in\mathcal{S}}\sum_{k\in\mathcal{K}_s} \theta_{s,k}\bigg\}.
\end{equation}
Further define the \textit{$\kappa$ percentile-spatial throughput} as the spatial throughput for the (disadvantaged) $\kappa$ percentile of UEs which have lower throughput in $\mathcal{K}$.
In this section, we use the 90\%-spatial throughput to measure the sum-rate performance for the disadvantaged majority of UEs.
Moreover, to measure the throughput fairness exactly, we adopt the Jain's fairness index.\footnote{Jain's fairness index $J$ for $\theta_k, k\in\mathcal{K}$, is defined as $J\triangleq \frac{(\mathbb{E}\{\theta\})^2}{\mathbb{E}\{\theta^2\}}$, where $J\in[0,1]$ and a higher $J$ represents better fairness.}
Finally, to measure the power consumption, we introduce the \textit{spatial transmit power (STP)} defined as the ratio of the average total TP versus the considered population area $S$, i.e., $\textrm{STP}\triangleq \frac{1}{S}\mathbb{E}\big\{\sum_{k\in\mathcal{K}} \delta_k P_k\big\}.$



\subsection{Single-GW Scenario}

\subsubsection{Verifying Throughput and Duty Cycle Formulas}
\begin{figure}[t]
	\centering
	\vspace{-1em}
	\includegraphics[width=1\linewidth,  trim=80 0 80 0,clip]{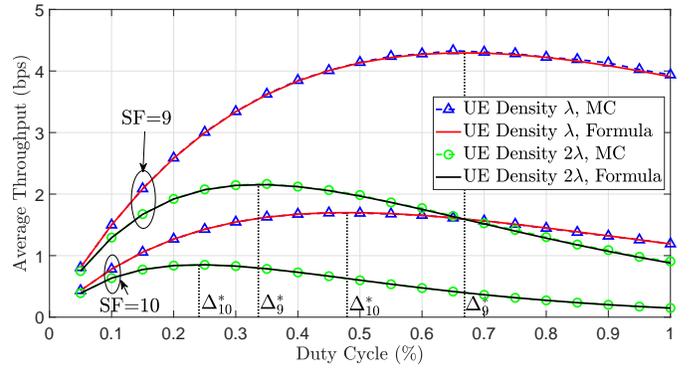}
	\caption{MC simulation verification for the throughput formula in \eqref{ThetasFinal} and the estimated optimal duty cycle in \eqref{DutyCycleOpt}.}\label{fig1}
	\vspace{-0.5em}
\end{figure}

In the first set of simulations, we verify the accuracy of the throughput formula in \eqref{ThetasFinal} and the estimated optimal duty cycle in \eqref{DutyCycleOpt}.
For the purpose of illustration, we consider a cell with radius $r_c=900$ m which is partitioned into six equal-width ring regions (i.e., $r_7=150$ m, $r_8=300$ m, $r_9=450$ m, etc.) each allocated with SF 7 to 12, respectively.
Under the slow channel inversion power control, we obtain the average (common) throughput of the UEs in each region for a given duty cycle, and plot the results for SF 9 and SF 10 in Fig. \ref{fig1},\footnote{Other SFs have similar results which are omitted for brevity.} under different UE density. 
It can be seen that the proposed formula matches quite well with the MC simulation results.
Moreover, under given UE density and SF allocation, it is observed that there exists an optimal duty cycle in maximizing the average throughput of the UEs with a given SF.
Therefore, it is beneficial for the network server to regulate the duty cycle level for traffic shaping in adaptation to the UE and/or GW densities, as is done in our proposed scheme in \eqref{DutyCycleOpt} and \eqref{DutyCycleOptMultiGW} for the single- and multi-GW cases, respectively.

\subsubsection{Impact of Discrete Power Levels}

\begin{figure}[t]
	\centering
	\vspace{-1em}
	\includegraphics[width=1.03\linewidth,  trim=50 0 30 0,clip]{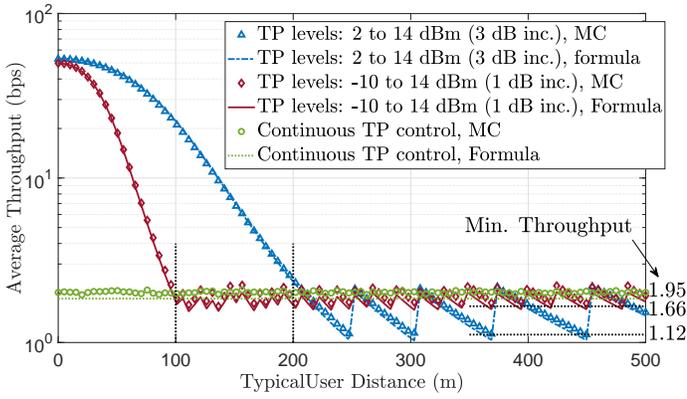}
	\caption{Impact of discrete power levels.}\label{TPdiscrete}
	\vspace{-0.5em}
\end{figure}

Here we investigate the impact of discrete TP levels on the throughput performance. For illustration, consider a single cell with radius $r_c=500$ m and all UEs transmitting with SF 7 and a duty cycle of 1$\%$. The throughput of the typical UE located at different distance from GW 0 is shown in Fig. \ref{TPdiscrete}, under three different TP profiles, respectively, i.e., 1) discrete TP levels: 2, 5, 8, 11 and 14 dBm (default by Semtech); 2) discrete TP levels: -10 to 14 dBm with 1 dB increment; and 3) continuous TP levels.

First, it can be seen that the generalized formula in \eqref{LaplaceDeriveDiscrete} to account for discrete power levels still matches well with the corresponding MC simulation results. Second, allowing more TP levels renders more flexibility and finer granularity for TP control, which helps improve both the minimum throughput and fairness index. 
For the three TP profiles above, the achieved minimum throughput is 1.12, 1.66 and 1.95 bps, with a fairness index of 0.206, 0.328 and 0.999, respectively.
Third, allowing a lower TP level for those UEs close to the GW not only saves more power, but also helps alleviate the near-far fairness issue. For the example in Fig. \ref{TPdiscrete}, the UEs using TP profile 1 within distance around 200 m from GW 0 (around $(200/500)^2=16\%$ of all UEs) transmit with 2 dBm power and have dominant throughput over others. In comparison, in the case with TP profile 2, the UEs with dominant throughput transmit with 0.1 mW power and reside within distance around 100 m from GW 0 (around $(100/500)^2=4\%$ of all UEs), while the common throughput for the majority of UEs has been improved from 1.12 bps to 1.66 bps. In the rest of simulations, we focus on the case with continuous TP levels in order to characterize the achievable common throughput upper bound.

\subsubsection{Achieving Max-Min Throughput}\label{SectionSimulationSingleMaxmin}
\begin{figure}[t]
	\centering
	\includegraphics[width=1\linewidth]{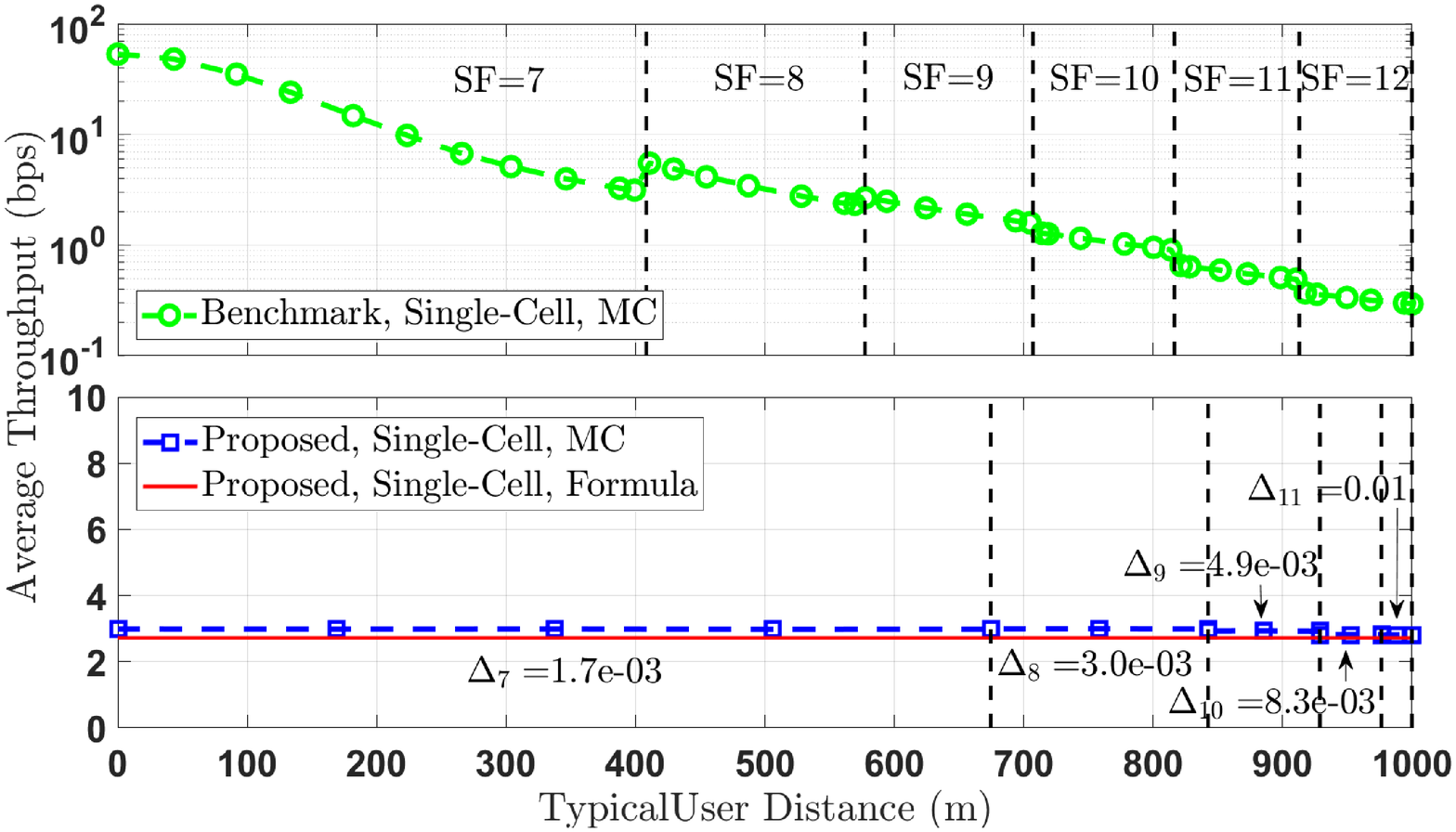}
	\caption{Throughput distribution in a single cell with radius 1 km.}\label{SingleCell1000m}
	\vspace{-1em}
\end{figure}

In this set of simulations, we apply our proposed scheme to achieve the average max-min throughput in a single-cell LoRa network, and also simulate a benchmark scheme using the MC method with fixed TP $P_k=P_{\textrm{max}}$ and fixed duty cycle $\delta_k=1\%$ for all UEs $k\in\mathcal{K}$, under the equal-area cell partitioning where the partitioning distance threshold $\bold r$ is set such that the ring regions associated with different SFs have the same area.
The results for a single cell with radius 1 km are plotted in Fig. \ref{SingleCell1000m}. For the benchmark scheme, it can be seen that the average throughput of the typical UE with a certain SF decreases with its distance from the GW, resulting in a prominent near-far fairness issue whereby the UEs close to the GW unfairly enjoy dominant throughput. Worse still, the UEs close to the cell edge (with SF 11 or 12) suffer from a poor throughput due to lower bit rate, resulting in a minimum throughput of $\bar\theta_\textrm{min}=0.29$ bps.
For our proposed scheme, it can be seen that our proposed formula in \eqref{ThetasFinal} for the throughput lower bound tightly matches the MC simulation results, and more UEs tend to be allocated with lower SF to enjoy higher bit rate.
In particular, the optimal duty cycle of SF 11 is capped by 1\%, while SF 12 is not used in this case due to its low bit rate.
As a result, the near-far fairness issue is greatly alleviated, and the throughput of cell-edge UEs greatly improved, with a minimum throughput of $\bar\theta_\textrm{min}=2.81$ bps per UE which is an order of magnitude higher than that of the benchmark scheme. 
Moreover, compared with the benchmark scheme, our proposed scheme improves the fairness index from 0.2145 to 0.9996, increases the 90\%-spatial throughput from 654.6 bps/km$^2$ to 930.5 bps/km$^2$, and also reduces the STP from 87.9 mW/km$^2$ to 22.8 mW/km$^2$ thanks to our proposed power control scheme.

\begin{figure}[t]
  \centering
  \includegraphics[width=1.0\linewidth,  trim=0 0 0 0,clip]{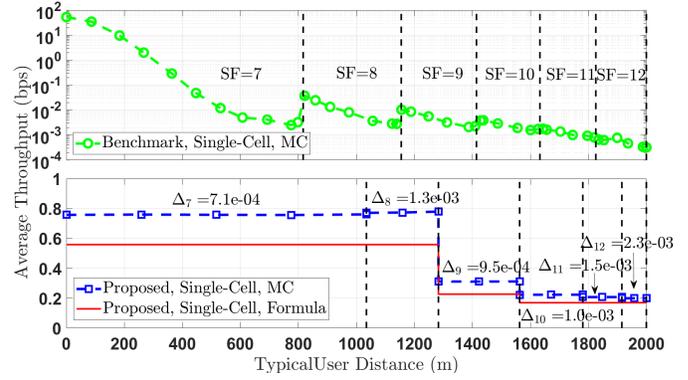}
  \caption{Throughput distribution in a single cell with radius 2 km.}\label{SingleCell2000m}
   \vspace{-1em}
\end{figure}

Next, we consider a larger cell with radius $r_c=2$ km and plotted the results in Fig. \ref{SingleCell2000m}. For the benchmark scheme, due to longer range and larger service area, it can be seen that the near-far fairness issue becomes more severe, and the throughput of cell-edge UEs becomes even poorer. 
In comparison, our proposed scheme greatly alleviates the near-far fairness issue and improves the cell-edge throughput. 
Moreover, compared with the benchmark scheme, our proposed scheme improves the fairness index from 0.0226 to 0.7614 and increases the 90\%-spatial throughput from 1.34 bps/km$^2$ to 134.4 bps/km$^2$, while the STP is further reduced from 87.9 mW/km$^2$ to 7.42 mW/km$^2$.
Finally, some deployment/design guidelines are suggested for a large service area. The properly-optimized LoRa network is able to serve the cell-edge UEs but typically at a low throughput, and it is beneficial to regulate the duty cycle level in adaptation to the network parameters including UE/GW densities so as to alleviate the collisions from massive co-SF devices.
This also motivates us to investigate the throughput fairness and scalability in the multi-GW scenarios, to further shed light on the network capacity or maximum supported UE density with rate requirements by densifying the GWs.

\subsection{Multi-GW Scenario}\label{SectionSimulationMultiCellSingleR}

In this subsection, we focus on the multi-GW LoRa networks under different cell sizes.
Since the interference from remote cells to the central cell is relatively weak, we consider a maximum interference range, denoted by $d_\textrm{max}$, whereby the cells within (or partially within) this range are taken into account. 
For the example of $d_\textrm{max}=3.2$ km, the number of considered cells is 7, 7, 13, 19 and 37 for the cases with cell radius $r_c=2.6$ km, 2 km, 1.5 km, 1 km and 700 m, respectively.
In the following, we first compare the performance of different frequency reuse schemes, and then focus on the 1-reuse scheme to illustrate how to achieve the average max-min throughput in the multi-GW scenario.

\subsubsection{Comparison of Different Frequency Reuse Schemes}\label{SectionSimulationReuse}
\begin{figure}[t]
	\centering
	\includegraphics[width=1.0\linewidth,  trim=50 0 70 0,clip]{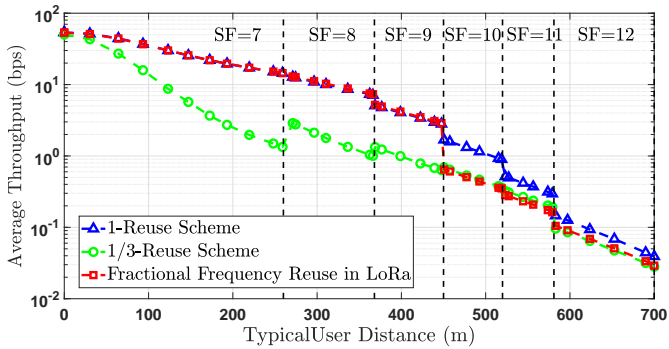}
	\caption{Comparison of different frequency reuse schemes.}\label{ReuseCurve}
	\vspace{-1em}
\end{figure}

For the purpose of exposition, we compare the 1-reuse, 1/3-reuse, and LoRa-FFR schemes in Section \ref{SectionMultiGW} under given TP and duty cycle as well as fixed SF allocation based on equal-area distance partitioning, similar to the setup of the benchmark scheme in Fig. \ref{SingleCell1000m} and Fig. \ref{SingleCell2000m}.
For fair comparison, here we assume that $\textrm{F}=3$ channels are available to serve UEs with density $\lambda_\textrm{active}=3\lambda$.
Consider $r_c=700$ m and 37 cells within range $d_\textrm{max}=3.2$ km. 
For the 1-reuse scheme, the UE density on each channel is $\lambda$ and the co-channel interference comes from all 37 cells.
For the 1/3-reuse scheme, the UE density on each channel is $3\lambda$ and the co-channel interference comes from 12 co-channel cells.
For the LoRa-FFR scheme, 
assume that the inner disk region associated with SFs 7, 8 and 9 use the 1-reuse scheme while the outer ring region associated with SFs 10, 11 and 12 use the 1/3-reuse scheme.
The results for single-GW reception is shown in Fig. \ref{ReuseCurve} while similar results for the multi-GW reception case are observed and hence omitted for brevity.

It can be seen that the 1-reuse scheme achieves overall higher throughput than that of the 1/3-reuse scheme, especially for the inner disk region close to GW 0. Due to the random (and hence non-orthogonal) time-sharing nature of Aloha, the co-channel interference comes from both cell 0 and other cells, whereby the intra-cell interference plays a more dominant role owing to its shorter distance from GW 0.
As a result, the 1-reuse scheme that thins the co-channel UE density in cell 0 is more effective in alleviating the co-channel interference level than the 1/3-reuse scheme which reduces interference from other cells.
The LoRa-FFR scheme can be treated as a hybrid scheme of the above two schemes and hence has corresponding performance on the respective regions.

Despite the achievable throughput performance, it is worth noting that the $1/\textrm{F}$-reuse and LoRa-FFR are still of value in practice, depending on the capability (and associated cost) of the LoRa GW chips used.
For example, Semtech SX1301 chip\cite{SX1301} can scan over multiple frequency channels, and supports a maximum of $N_D=8$ concurrent demodulation paths for arbitrary combination of operating channel and SF. Note that $N_D$ can be set to any value for a customer specific circuit with the corresponding number of demodulation circuits installed.
Therefore, since the $1/\textrm{F}$-reuse and LoRa-FFR schemes may require smaller $N_D$ compared with the 1-reuse scheme, they provide different options for capacity and cost trade-off.
In the rest, we focus on the 1-reuse scheme on a single channel.

\subsubsection{Achieving Max-Min Throughput}\label{SectionSimulationMultiSingle2}

\begin{figure}[t]
	\centering
	\includegraphics[width=1.0\linewidth]{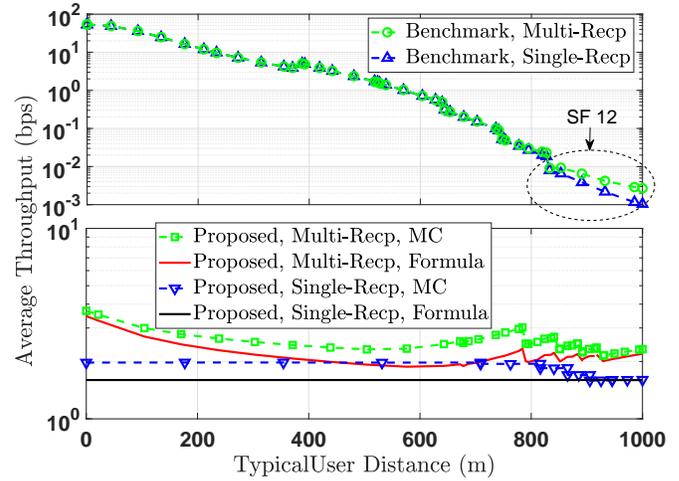}
	\caption{Throughput distribution in cell 0 for the multi-GW scenario with $r_c=1$ km.}\label{Joint1000m}
	\vspace{-1em}
\end{figure}

In this subsection, we compare the performance of our proposed scheme with the benchmark scheme (similar to the benchmark setup in Fig. \ref{SingleCell1000m}) under both single-GW reception and multi-GW reception.
For illustration, consider the multi-GW scenario with cell radius $r_c=1$ km. The obtained results are shown in Fig. \ref{Joint1000m}.

Similar to the results in Fig. \ref{SingleCell1000m} for the single-cell case, our proposed scheme greatly improves the achieved throughput and fairness.
Under single-GW reception, compared with the benchmark scheme, our proposed scheme improves the fairness index from 0.1477 to 0.9975, increases the minimum throughput from 0.001 bps to 1.591 bps, increases the 90\%-spatial throughput from 395.1 bps/km$^2$ to 605.0 bps/km$^2$, and also reduces the STP from 87.9 mW/km$^2$ to 11.5 mW/km$^2$.

In the multi-GW reception scenario, it can be seen from Fig. \ref{Joint1000m} that the benchmark scheme has almost no improvement over the single-GW reception counterpart, except marginal improvement for UEs with SF 12 which locate at the cell edge and are closer to other neighboring GWs.
This is due to the lack of power control in the benchmark scheme which results in high aggregate interference power at the GWs and hence reduces the multi-GW reception probability.
In contrast, thanks to our joint power control and duty cycle adjustment, our proposed scheme reduces the aggregate interference at the GWs and hence improves the throughput for all the UEs with different SFs under multi-GW reception.
As a result, in the multi-GW reception scenario, the minimum throughput is improved to 2.147 bps and the 90\%-spatial throughput is improved to 779.3 bps/km$^2$, with comparable fairness index of 0.9914 and STP of 12.8 mW/km$^2$.

\subsection{Fairness and Scalability under Different UE/GW Densities}\label{SectionSimulationScalability}

In this subsection, we investigate the network fairness and scalability under different UE/GW densities, where the results for the minimum throughput, 90\%-spatial throughput, throughput fairness index and spatial transmit power (STP) are plotted in Figures \ref{MinThroughputCurve} to \ref{STPCurve}, respectively.

\subsubsection{Minimum Throughput}\label{SectionMinThr}
\begin{figure}[t]
	\centering
	\includegraphics[width=0.9\linewidth,  trim=40 0 50 0,clip]{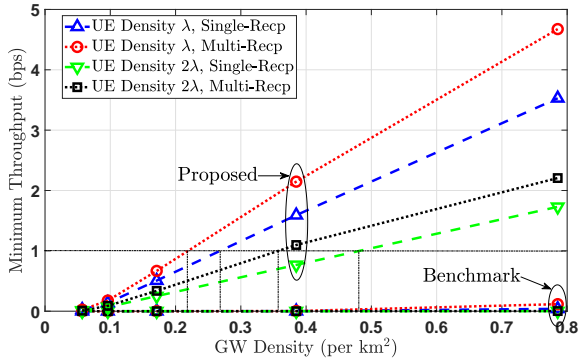}
	\caption{Minimum throughput under different UE/GW densities.}\label{MinThroughputCurve}
	\vspace{-1em}
\end{figure}

It can be seen from Fig. \ref{MinThroughputCurve} that our proposed scheme significantly improves the minimum throughput as the GW density increases or as the UE density decreases, and is able to effectively exploit the multi-GW reception diversity in achieving higher throughput.
In comparison, the benchmark scheme suffers from severe fairness issue discussed in the above subsections, whereby the UEs close to the receiving GW enjoy dominant throughput performance while the cell-edge UEs have very poor throughput.
As a result, the minimum throughput of the benchmark scheme remains very low despite the increase of GW density.

Moreover, the results in Fig. \ref{MinThroughputCurve} provide quick reference for the maximum supported UE density with rate requirements under given GW density and other network parameters, or equivalently, the required GW density to satisfy such rate requirements.
For the example setup considered and a target rate requirement of 1 bps per UE, the required GW density to support a UE density of $\lambda=350$/km$^2$ (or $2\lambda=700$/km$^2$) per channel of 125 kHz, is around 0.27 and 0.22 GWs/km$^2$ (or around 0.48 and 0.36 GWs/km$^2$) for the cases with single- and multi-GW reception, respectively.

\subsubsection{90\%-Spatial Throughput}

\begin{figure}[t]
	\centering
	\includegraphics[width=0.9\linewidth,  trim=40 0 70 0,clip]{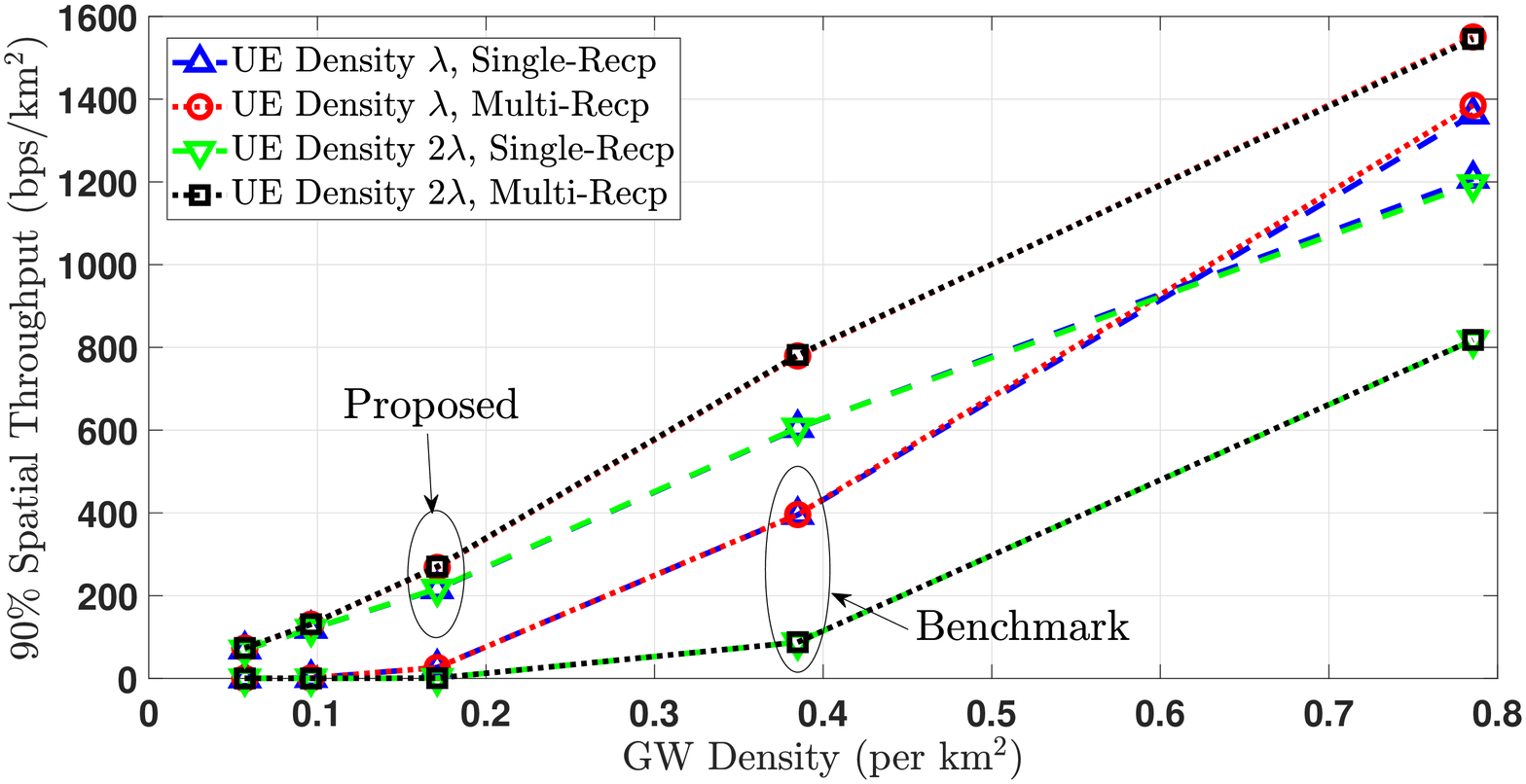}
	\caption{90\%-spatial throughput under different UE/GW densities.}\label{SpatialThroughputCurve}
	\vspace{-1em}
\end{figure}

It can be seen from Fig. \ref{SpatialThroughputCurve} that the 90\%-spatial throughput increases as the GW density increases, for both the benchmark scheme and our proposed scheme.
However, such improvement comes mostly from the cell-center UEs (dominant minority) in the benchmark scheme, while our proposed scheme fairly improves the throughput of all UEs.
As a result, as the UE density increases, the 90\%-spatial throughput of the benchmark scheme drops significantly since most throughput is reaped by the cell-center UEs, while that of our proposed scheme remains almost the same thanks to our joint optimization algorithm that is able to adapt to different network size and UE density.
Moreover, our proposed scheme can better exploit the multi-GW reception diversity than the benchmark scheme.

\subsubsection{Throughput Fairness Index}

\begin{figure}[t]
	\centering
	\includegraphics[width=0.9\linewidth,  trim=50 0 70 0,clip]{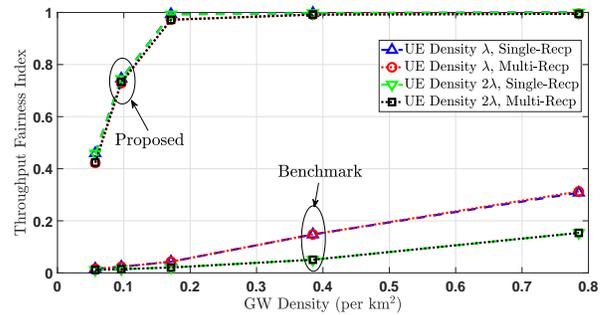}
	\caption{Throughput fairness index under different UE/GW densities.}\label{FairnessCurve}
	\vspace{-1em}
\end{figure}

It can be seen from Fig. \ref{FairnessCurve} that the throughput fairness index increases as the GW density increases for both the benchmark scheme and our proposed scheme, since the corresponding cell radius becomes smaller and the near-far fairness issue becomes less severe.
However, the fairness index of the benchmark scheme remains at a low level and decreases as the UE density increases, due to similar reasons discussed in Section \ref{SectionMinThr}.
In contrast, our proposed scheme adapts to the UE/GW densities and achieves a high level of throughput fairness under both single-GW/multi-GW reception.

\subsubsection{Spatial Transmit Power (STP)}

\begin{figure}[t]
	\centering
	\includegraphics[width=0.9\linewidth,  trim=50 0 70 0,clip]{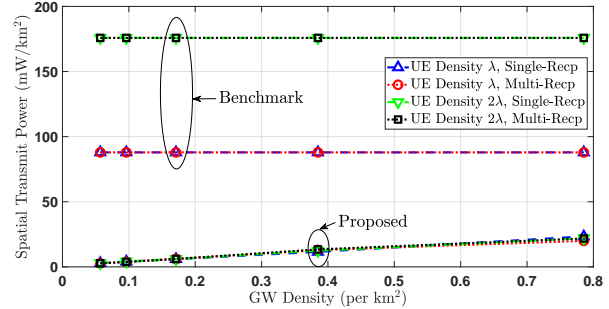}
	\caption{Spatial transmit power under different UE/GW densities.}\label{STPCurve}
	\vspace{-1em}
\end{figure}

It can be seen from Fig. \ref{STPCurve} that the STP of the benchmark scheme is high and also increases linearly as the UE density increases.
In comparison, our proposed scheme adapts to the UE/GW densities by joint power control and duty cycle adjustment, and hence is able to save the STP significantly regardless of the UE density.

In summary, compared with the benchmark scheme, our proposed scheme with joint power control, duty cycle adjustment and SF allocation not only alleviates the near-far fairness issue, but also significantly improves the minimum throughput and 90\%-spatial throughput as well as saving the STP, regardless of the UE/GW densities.
Moreover, our proposed scheme is better at exploiting the multi-GW reception diversity in achieving higher throughput improvement.
Finally, the results in Section \ref{SectionSimulationScalability} provides a good reference for network operators in deploying LoRa GWs/UEs in the considered area with a proper density, in order to satisfy the overall throughput fairness and scalability requirement in the large-scale LoRa network.

\section{Conclusions}\label{SectionConclusion}
To achieve massive connectivity with fairness in large-scale LoRa networks, 
we leverage on stochastic geometry, especially the Poisson rain model, and derive (semi-) closed-form formulas for the aggregate interference distribution, packet success probability and hence system throughput in both single-cell and multi-cell setups with frequency reuse, by accounting for channel fading, random UE distribution, partial packet overlapping, and/or multi-GW packet reception.
Based on the derived analytical formulas which require only average channel statistics and spatial UE distribution, we further propose an IB method that quickly yields high-level policies of joint SF/TP/duty cycle control in adaptation to network parameters including UE/GW densities, for gauging the average max-min UE throughput or supported UE density with rate requirements.
Numerical results validate the analytical formulas and the effectiveness of our proposed optimization scheme, which greatly alleviates the near-far fairness issue and reduces the spatial power consumption, while significantly improving the cell-edge throughput as well as the spatial (sum) throughput for the majority of UEs.
Finally, it is shown that the macro diversity of multi-GW reception can be effectively exploited by our proposed scheme in achieving higher throughput improvement.
Future work could further consider confirmed traffic and incorporate downlink interference into the interference characterization.


\bibliography{IEEEabrv,BibDIRP}

\begin{thebibliography}{10}
\providecommand{\url}[1]{#1}
\csname url@samestyle\endcsname
\providecommand{\newblock}{\relax}
\providecommand{\bibinfo}[2]{#2}
\providecommand{\BIBentrySTDinterwordspacing}{\spaceskip=0pt\relax}
\providecommand{\BIBentryALTinterwordstretchfactor}{4}
\providecommand{\BIBentryALTinterwordspacing}{\spaceskip=\fontdimen2\font plus
\BIBentryALTinterwordstretchfactor\fontdimen3\font minus
  \fontdimen4\font\relax}
\providecommand{\BIBforeignlanguage}[2]{{%
\expandafter\ifx\csname l@#1\endcsname\relax
\typeout{** WARNING: IEEEtran.bst: No hyphenation pattern has been}%
\typeout{** loaded for the language `#1'. Using the pattern for}%
\typeout{** the default language instead.}%
\else
\language=\csname l@#1\endcsname
\fi
#2}}
\providecommand{\BIBdecl}{\relax}
\BIBdecl

\bibitem{LoRaMaxMinArXiv}
J.~{Lyu}, D.~{Yu}, and L.~{Fu}, ``Achieving max-min throughput in {LoRa}
  networks,'' in \emph{Int. Conf. Comput., Netw. and Commun. (ICNC)}, 2020, pp.
  471--476.

\bibitem{LPWANoverview}
U.~{Raza}, P.~{Kulkarni}, and M.~{Sooriyabandara}, ``Low power wide area
  networks: {An} overview,'' \emph{IEEE Commun. Surveys Tuts.}, vol.~19, no.~2,
  pp. 855--873, 2017.

\bibitem{NBiotMagazine2016}
J.~{Gozalvez}, ``New {3GPP} standard for {IoT} [mobile radio],'' \emph{IEEE
  Veh. Technol. Mag.}, vol.~11, no.~1, pp. 14--20, Mar. 2016.

\bibitem{LoRaWAN}
{\relax LoRa Alliance Technical Committee}, \emph{{LoRaWAN 1.0.4 Specification
  Package}}.\hskip 1em plus 0.5em minus 0.4em\relax LoRaWAN Alliance, Oct.
  2020.

\bibitem{SemTechLoRaModulationBasics}
{\relax Semtech}, \emph{{LoRa Modulation Basics, AN1200.22 Revision 2}}.\hskip
  1em plus 0.5em minus 0.4em\relax Semtech, May 2015.

\bibitem{LoRaSurvey}
J.~P. {Shanmuga Sundaram}, W.~{Du}, and Z.~{Zhao}, ``A survey on {LoRa}
  networking: {Research} problems, current solutions, and open issues,''
  \emph{IEEE Commun. Surveys Tuts.}, vol.~22, no.~1, pp. 371--388, 2020.

\bibitem{LoRaIoTJ2018Pollin}
B.~{Reynders}, Q.~{Wang}, P.~{Tuset-Peiro}, X.~{Vilajosana}, and S.~{Pollin},
  ``Improving reliability and scalability of {LoRaWANs} through lightweight
  scheduling,'' \emph{IEEE Internet Things J.}, vol.~5, no.~3, pp. 1830--1842,
  Jun. 2018.

\bibitem{LoRaIoTJ2019LowOverhead}
J.~{Haxhibeqiri}, I.~{Moerman}, and J.~{Hoebeke}, ``Low overhead scheduling of
  {LoRa} transmissions for improved scalability,'' \emph{IEEE Internet Things
  J.}, pp. 1--1, 2019.

\bibitem{LoRaSensors2017}
J.~Haxhibeqiri, F.~Van~den Abeele, I.~Moerman, and J.~Hoebeke, ``{LoRa
  scalability: A simulation model based on interference measurements},''
  \emph{Sensors}, vol.~17, no.~6, p. 1193, 2017.

\bibitem{LoRaNS3IoTJ}
F.~{Van den Abeele}, J.~{Haxhibeqiri}, I.~{Moerman}, and J.~{Hoebeke},
  ``Scalability analysis of large-scale {LoRaWAN} networks in ns-3,''
  \emph{IEEE Internet Things J.}, vol.~4, no.~6, pp. 2186--2198, Dec. 2017.

\bibitem{SemtechDenseTrail}
{\relax Semtech}, \emph{{LoRaWAN’s capacity trials in dense urban
  environments}}.\hskip 1em plus 0.5em minus 0.4em\relax Semtech, 2018.

\bibitem{AndrewsCellular}
J.~G. Andrews, F.~Baccelli, and R.~K. Ganti, ``A tractable approach to coverage
  and rate in cellular networks,'' \emph{IEEE Trans. Commun.}, vol.~59, no.~11,
  pp. 3122--3134, Nov. 2011.

\bibitem{SpatialAlohaSlotted}
F.~{Baccelli}, B.~{Blaszczyszyn}, and P.~{Muhlethaler}, ``An {Aloha} protocol
  for multihop mobile wireless networks,'' \emph{IEEE Trans. Inf. Theory},
  vol.~52, no.~2, pp. 421--436, Feb. 2006.

\bibitem{LoRaWCL2017Raza}
O.~Georgiou and U.~Raza, ``Low power wide area network analysis: {Can LoRa}
  scale?'' \emph{IEEE Wireless Commun. Lett.}, vol.~6, no.~2, pp. 162--165,
  Apr. 2017.

\bibitem{MahmoodInterSF}
A.~{Mahmood}, E.~{Sisinni}, L.~{Guntupalli}, R.~{Rondón}, S.~A. {Hassan}, and
  M.~{Gidlund}, ``Scalability analysis of a {LoRa} network under imperfect
  orthogonality,'' \emph{IEEE Trans. Industr. Inform.}, vol.~15, no.~3, pp.
  1425--1436, 2019.

\bibitem{LoRaCL2018Korean}
J.~Lim and Y.~Han, ``Spreading factor allocation for massive connectivity in
  {LoRa} systems,'' \emph{IEEE Commun. Lett.}, vol.~22, no.~4, pp. 800--803,
  Apr. 2018.

\bibitem{2DinterferenceICC2017}
Z.~{Li}, S.~{Zozor}, J.~{Brossier}, N.~{Varsier}, and Q.~{Lampin}, ``{2D}
  time-frequency interference modelling using stochastic geometry for
  performance evaluation in low-power wide-area networks,'' in \emph{Proc. IEEE
  Int. Conf. Commun.}, May 2017, pp. 1--7.

\bibitem{NonSlottedINFOCOM2010}
B.~Blaszczyszyn and P.~Muhlethaler, ``Stochastic analysis of non-slotted
  {Aloha} in wireless ad-hoc networks,'' in \emph{Proc. IEEE INFOCOM}, Mar.
  2010, pp. 1--9.

\bibitem{PoissonRainLoRa}
B.~B\l{}aszczyszyn and P.~M\"{u}hlethaler, ``Analyzing {LoRa} long-range,
  low-power, wide-area networks using stochastic geometry,'' in \emph{EAI Int.
  Conf. Performance Evaluation Methodologies and Tools}, 2019, p. 119–126.

\bibitem{SemTechADR}
{\relax Semtech}, \emph{{LoRaWAN - Simple Rate Adaptation Recommended
  Algorithm}}.\hskip 1em plus 0.5em minus 0.4em\relax Semtech, Oct. 2016.

\bibitem{RazaADR}
S.~{Li}, U.~{Raza}, and A.~{Khan}, ``How agile is the adaptive data rate
  mechanism of {LoRaWAN}?'' in \emph{Proc. IEEE GLOBECOM}, 2018, pp. 206--212.

\bibitem{ADRsurvey}
R.~Kufakunesu, G.~P. Hancke, and A.~M. Abu-Mahfouz, ``A survey on adaptive data
  rate optimization in {LoRaWAN: Recent} solutions and major challenges,''
  \emph{Sensors}, vol.~20, no.~18, 2020.

\bibitem{BianchiLoRaTWC}
D.~{Garlisi}, I.~{Tinnirello}, G.~{Bianchi}, and F.~{Cuomo}, ``Capture aware
  sequential waterfilling for {LoRaWAN} adaptive data rate,'' \emph{IEEE Trans.
  Wireless Commun.}, to appear.

\bibitem{LoRaPollinICC2017}
B.~{Reynders}, W.~{Meert}, and S.~{Pollin}, ``Power and spreading factor
  control in low power wide area networks,'' in \emph{Proc. IEEE Int. Conf.
  Commun.}, May 2017, pp. 1--6.

\bibitem{HeusseSpatial}
A.~Duda and M.~Heusse, ``Spatial issues in modeling {LoRaWAN} capacity,'' in
  \emph{Proc. ACM Int. Conf. MSWiM}, 2019, p. 191–198.

\bibitem{QinZhijinEE}
B.~{Su}, Z.~{Qin}, and Q.~{Ni}, ``Energy efficient uplink transmissions in
  {LoRa} networks,'' \emph{IEEE Trans. Commun.}, vol.~68, no.~8, pp.
  4960--4972, 2020.

\bibitem{JimingLoRa}
Y.~{Sun}, J.~{Chen}, S.~{He}, and Z.~{Shi}, ``High-confidence gateway planning
  and performance evaluation of a hybrid {LoRa} network,'' \emph{IEEE Internet
  Things J.}, vol.~8, no.~2, pp. 1071--1081, 2021.

\bibitem{AndrewsPrimer}
\BIBentryALTinterwordspacing
J.~G. Andrews, A.~K. Gupta, and H.~S. Dhillon, ``A primer on cellular network
  analysis using stochastic geometry,'' 2016. [Online]. Available:
  \url{https://arxiv.org/abs/1604.03183}
\BIBentrySTDinterwordspacing

\bibitem{AndrewsGeneralFading}
R.~{Tanbourgi}, H.~S. {Dhillon}, J.~G. {Andrews}, and F.~K. {Jondral},
  ``Dual-branch {MRC} receivers under spatial interference correlation and
  {Nakagami} fading,'' \emph{IEEE Trans. Commun.}, vol.~62, no.~6, pp.
  1830--1844, June 2014.

\bibitem{LoRaThreshold2015}
C.~Goursaud and J.-M. Gorce, ``{Dedicated networks for {IoT: PHY / MAC} state
  of the art and challenges},'' \emph{{EAI endorsed trans. Internet of
  Things}}, Oct. 2015.

\bibitem{CroceCLinterSF}
D.~{Croce}, M.~{Gucciardo}, S.~{Mangione}, G.~{Santaromita}, and
  I.~{Tinnirello}, ``Impact of {LoRa} imperfect orthogonality: {Analysis} of
  link-level performance,'' \emph{IEEE Commun. Lett.}, vol.~22, no.~4, pp.
  796--799, 2018.

\bibitem{HeusseSingleCell}
M.~Heusse, T.~Attia, C.~Caillouet, F.~Rousseau, and A.~Duda, ``Capacity of a
  {LoRaWAN} cell,'' in \emph{Proc. ACM Int. Conf. MSWiM}, 2020, pp. 131--140.

\bibitem{LoRaMacroDiversity}
Q.~{Song}, X.~{Lagrange}, and L.~{Nuaymi}, ``Evaluation of macro diversity gain
  in long range {ALOHA} networks,'' \emph{IEEE Commun. Lett.}, vol.~21, no.~11,
  pp. 2472--2475, Nov. 2017.

\bibitem{SX1301}
\BIBentryALTinterwordspacing
{\relax Semtech}, \emph{{SX1301 Datasheet}}. [Online]. Available:
  \url{https://www.semtech.com/products/wireless-rf/lora-gateways/sx1301}
\BIBentrySTDinterwordspacing

\end{thebibliography}

\newpage

\end{document}